\documentclass[11pt]{article}
\usepackage[utf8]{inputenc}
\usepackage[margin=1in]{geometry}
\usepackage{amsmath,amssymb}
\allowdisplaybreaks[1]
\usepackage{graphicx}
\usepackage{hyperref}
\usepackage{xcolor}
\usepackage{setspace} 
\usepackage[superscript,biblabel]{cite}
\usepackage{booktabs,longtable,tabularx,array,makecell}
\usepackage{caption}
\usepackage{placeins}
\usepackage{float}
\newcolumntype{Y}{>{\raggedright\arraybackslash}X}

\renewenvironment{abstract}
 {\small
  \begin{center}
    \bfseries\abstractname\vspace{-0.5em}\vspace{0pt}
  \end{center}
  \list{}{%
    \setlength{\leftmargin}{0pt}%
    \setlength{\rightmargin}{0pt}%
  }%
  \item\relax}
 {\endlist}

\definecolor{titleblue}{RGB}{0, 0, 0}
 
\title{\color{titleblue}\textbf{Robust Decision Curve Analysis for Survival Risk Prediction Models in the Presence of Informative Censoring}}
 
\author{
\resizebox{0.98\textwidth}{!}{%
Amirhosein Farahmand$^{1}$, Mauricio Lopez-Mendez$^{2}$, Mohsen Sadatsafavi$^{3}$, Abdollah Safari$^{1,*,\dagger}$, Tae Yoon Lee$^{4,*,\dagger}$%
}\\[0.5em]
{\small \textbf{Affiliations:}}\\[0.2em]
{\small $^{1}$School of Mathematics, Statistics, and Computer Science, College of Science,}\\
{\small University of Tehran, Tehran, Iran}\\[0.1em]
{\small $^{2}$Department of Health Policy, Stanford School of Medicine, Stanford University,}\\
{\small Stanford, CA, USA}\\[0.1em]
{\small $^{3}$Respiratory Evaluation Sciences Program, Faculty of Pharmaceutical Sciences,}\\
{\small The University of British Columbia, Vancouver, Canada}\\[0.1em]
{\small $^{4}$Quantitative Sciences Unit, Division of Computational Medicine, Stanford}\\
{\small University School of Medicine, Palo Alto, CA, USA}
}
 
\date{}

\begin{document}
\onehalfspacing %
 
\maketitle
\vspace{-3.3em}
\begingroup
\renewcommand{\thefootnote}{}
\footnotetext{$^{*}$Equal contribution.}
\footnotetext{$^{\dagger}$Corresponding authors: Abdollah Safari (\href{mailto:a.safari@ut.ac.ir}{\nolinkurl{a.safari@ut.ac.ir}}), University of Tehran; Tae Yoon Lee (\href{mailto:harrytyl@stanford.edu}{\nolinkurl{harrytyl@stanford.edu}}), Stanford University School of Medicine.}
\endgroup

\begin{abstract}
\singlespacing
Decision curve analysis (DCA) is recommended for evaluating the clinical utility of prediction models by quantifying net benefit (NB). Conventional NB estimators for survival outcomes rely on the assumption that censoring is non-informative, potentially yielding biased estimates and misleading decisions when this assumption is violated. We propose a robust estimator of survival NB using conditional inverse probability of censoring weighting (CIPCW). In simulation studies, we compared its performance against the conventional approach across censoring mechanisms, censoring model specifications, censoring rates, and decision thresholds, in terms of relative bias, relative root mean squared error (RMSE), and 90\% confidence interval coverage. Under non-informative censoring, both estimators performed comparably, with relative bias below 1\% and coverage close to the nominal level. Under informative censoring, conventional estimators underestimated the true NB by up to 58\% with coverage as low as 1\%, whereas the CIPCW estimator had a relative bias of at most 4\%, coverage of at least 78\% and consistently lower relative RMSE. As a case study, we illustrated its use in validating a breast cancer recurrence-free survival model to guide intensified adjuvant therapy in the German Breast Cancer Study Group (GBSG) cohort. Using the model was the best strategy at some clinically relevant thresholds, but accounting for potential informative censoring changed the best strategy to treating all patients, showing that informative censoring can change the conclusions drawn from DCA. When censoring may be informative, we recommend CIPCW-based NB estimation in DCA for survival outcomes.
\newpage


\end{abstract}

\textbf{Highlights}
\begin{itemize}
    \item Conventional decision curve analysis for survival outcomes relies on the assumption of non-informative censoring, which may be violated when follow-up depends on patient characteristics associated with the event time of interest.
    
    \item Under informative censoring, net benefit estimators based on conventional survival methods can be biased and lead to suboptimal clinical decisions. 
    
    \item We developed a robust estimator of net benefit based on conditional inverse probability of censoring weighting and demonstrated its robustness to informative censoring through a Monte Carlo simulation study and illustrated its utility in a case study of recurrence-free survival in women with breast cancer.
    
\end{itemize}

\newpage
\section{Introduction}
 
Before a risk prediction model is used to guide treatment decisions, it should be validated in the population in which it is intended to be used.\cite{binuya2022methodological} Net benefit (NB) is a widely used metric for assessing the clinical utility of such a model\cite{vickers2006dca}, recommended by best practice guidelines\cite{collins2024tripodai, vancalster2025ai}. It is computed at a decision threshold that turns the predicted risk into a binary action (treat or not treat), whereby treatment is offered to patients whose predicted risk exceeds this threshold. NB weighs the benefit of treating an individual who would experience the event against the harm of unnecessarily treating one who would not. The threshold is chosen a priori based on the clinical context~\cite{wynants2019myths} and sets the exchange rate between this benefit and harm~\cite{vickers2006dca, vickers2016net}. Because this exchange rate is usually not known exactly, NB is often calculated and plotted across a clinically relevant range of thresholds, referred to as decision curve analysis (DCA)~\cite{vickers2006dca, fitzgerald2015dca}. Following the landmark paper on NB for binary outcomes\cite{vickers2006dca}, Vickers et al. extended NB to survival outcomes using the Kaplan-Meier (KM) and Aalen-Johansen (AJ) approaches in the absence and presence of competing risks, respectively, to account for potential censoring~\cite{vickers2008extensions}.
 
However, both approaches assume non-informative censoring,~\cite{heagerty2000time, collett2023modelling} an assumption that often fails in practice. When the factors that drive loss to follow-up, such as disease severity or specific biomarkers, also influence outcome risk, patients who remain under observation no longer represent the full cohort~\cite{lin2023informative,robins2000correcting, altawil2024addressing}. For example, in a prostate cancer study investigating the prediction of response to radiotherapy,~\cite{gerds2013estimating} censoring was strongly associated with radiation dose, a factor that also affects treatment response, necessitating adjustment for informative censoring to obtain an unbiased concordance index. In a follow-up simulation study based on these data, the risk of this dependency was confirmed, showing that censoring models which omitted radiation dose or assumed independence introduced substantial bias into the estimates of model performance.~\cite{gerds2013estimating} Under informative censoring, standard NB calculations may therefore misrepresent a model's clinical utility, and a model could be adopted or rejected for the wrong reasons.
 
In this work, we propose an estimator based on conditional inverse probability of censoring weighting (CIPCW) that stays valid under informative censoring, with or without competing risks. In simulations across a range of censoring mechanisms and model specifications, we investigate its performance against the conventional estimator. Applying this method to publicly available clinical trial data, we show that accounting for potential informative censoring can change the conclusions drawn from DCA. We implement the CIPCW-based estimator in a fork of the \texttt{dcurves} R package, so researchers can run robust DCA for survival outcomes directly.

\section{Method}

\subsection{Setting and Notations}

Given a validation dataset, DCA evaluates the clinical utility of a prespecified survival prediction model by considering decision rules of the form ``treat if predicted risk exceeds a threshold $z$'' across a range of clinically relevant thresholds.

Let $T$ denote the time to event, $C$ the censoring time, and $X \in \mathbb{R}^p$ a $p$-dimensional vector of baseline covariates. The cause of the event is denoted by $J \in \{1, \ldots, M\}$. The observed time is defined as $\widetilde{T} = \min(T, C)$, and the observed cause indicator $\Delta = I(T \leq C) J \in \{0, 1, \ldots, M\}$ specifies the cause of the event when $\Delta \neq 0$, or indicates censoring when $\Delta = 0$. We observe an independent and identically distributed (i.i.d.) sample of $n$ subjects, denoted by $\{(\widetilde{T}_i, \Delta_i, X_i)\}_{i=1,\ldots,n}$. We consider a survival prediction model developed using an independent training dataset, which for a suitably selected prediction horizon $\tau$, provides the predicted probability $\pi^{(k)} \equiv \pi^{(k)}(X, \tau)$, mapping the covariate pattern $X$ to the predicted risk of experiencing cause $k$ by time $\tau$. For subject $i$, this predicted probability is denoted $\pi_i^{(k)} = \pi^{(k)}(X_i, \tau)$. Throughout, we refer to the event of interest as cause $k$, and the standard survival setting is the special case $M = 1$, in which there is only one cause and it is the event of interest.

\subsection{Net Benefit for Survival Outcomes}

For a binary outcome, NB at a decision threshold $z$ can be expressed in terms of outcome prevalence (Prev), sensitivity (Se), and specificity (Sp)\cite{wynants2018random}:

\begin{equation}
\mathrm{NB}
=
\mathrm{Prev}\cdot\mathrm{Se}
-
(1-\mathrm{Prev})\cdot(1-\mathrm{Sp})\cdot
\frac{z}{1-z}.
\label{eq:nb}
\end{equation}
For a survival outcome, these quantities must be defined at the prediction horizon $\tau$ at which the decision is evaluated. At prediction horizon \(\tau\) and decision threshold \(z\), sensitivity is the probability of a predicted risk above \(z\) among individuals who experience cause \(k\) by \(\tau\), whereas specificity is the probability of a predicted risk at or below \(z\) among those who remain event-free at \(\tau\) or experience a competing event by \(\tau\).\cite{kamarudin2017time, blanche2013estimating} For simplicity, we refer to these quantities as time-dependent sensitivity ($\mathrm{Se}_\tau^{(k)}$) and specificity ($\mathrm{Sp}_\tau^{(k)}$), respectively. Additionally, the time-dependent prevalence ($\mathrm{Prev}_\tau^{(k)}$) is defined as the marginal probability of experiencing cause \(k\) by \(\tau\). At the population level, these quantities are defined as:

$$
\begin{gathered}
\mathrm{Prev}_{\tau}^{(k)} = P(T\leq\tau,J=k), \qquad
\mathrm{Se}_{\tau}^{(k)} = P(\pi^{(k)}>z\mid T\leq\tau,J=k), \\[1ex]
\mathrm{Sp}_{\tau}^{(k)} = P\!\left(\pi^{(k)}\leq z\mid T>\tau\ \text{or}\ (T\leq\tau,J\neq k)\right).
\end{gathered}
$$
When $M = 1$, the competing-event set is empty, so $\mathrm{Prev}_{\tau}^{(k)}$ and $\mathrm{Se}_{\tau}^{(k)}$ are simplified by removing the condition \(J=k\), and $\mathrm{Sp}_{\tau}^{(k)}$ is simplified by removing the term \(\{T\leq\tau,\,J\neq k\}\), yielding the time-dependent prevalence, sensitivity, and specificity for a standard survival outcome.

Substituting $\mathrm{Prev}_{\tau}^{(k)}$, $\mathrm{Se}_{\tau}^{(k)}$, and $\mathrm{Sp}_{\tau}^{(k)}$ into Equation~\eqref{eq:nb} gives the NB for cause \(k\) at prediction horizon \(\tau\) and decision threshold \(z\):

\begin{equation}
\mathrm{NB}_{\tau}^{(k)}
=
\mathrm{Prev}_{\tau}^{(k)}\cdot\mathrm{Se}_{\tau}^{(k)}
-
(1-\mathrm{Prev}_{\tau}^{(k)})\cdot(1-\mathrm{Sp}_{\tau}^{(k)})\cdot
\frac{z}{1-z}.
\label{eq:survival_nb}
\end{equation}
Under the non-informative censoring assumption, Vickers et al.\cite{vickers2008extensions} proposed consistent estimators of survival NB using the KM and AJ approaches for the standard survival and competing-risks settings, respectively.

\subsection{The Robust CIPCW-Based Estimator}

The conventional KM and AJ-based estimators can become biased when their non-informative censoring assumption is violated.~\cite{collett2023modelling, kamarudin2017time} To address the issues with potential informative censoring, we propose a robust estimator for survival NB based on CIPCW methodology, drawing on the general framework for censored and missing data~\cite{vanderlaan2003unified} and related applications~\cite{gerds2006consistent,gerds2013estimating,prince2025estimation,blanche2013estimating}.

Let $G(u\mid\mathbf{X}_i)=P(C_i>u\mid\mathbf{X}_i)$ denote the conditional survival function of the censoring time, with \(\widehat{G}(u\mid\mathbf{X}_i)\) being its estimate. This estimate can be obtained using an appropriate survival model by treating censoring as the event, according to the complexity of the underlying censoring mechanism. Accordingly, individuals who experience any event at \(\widetilde{T}_i\) are weighted by \(W_i = \widehat{G}(\widetilde{T}_i\mid\mathbf{X}_i)^{-1}\), whereas those known to remain event-free through \(\tau\) are weighted by \(W_i^\prime = \widehat{G}(\tau\mid\mathbf{X}_i)^{-1}\). Through this reweighting, CIPCW creates a pseudo-population in which the observed contributions are weighted to represent individuals with similar covariates whose status at $\tau$ is unobserved because of censoring, thereby correcting the selection induced by informative censoring.

For this pseudo-population to represent the target population, we assume throughout that: (1) censoring time is conditionally independent of the event time and cause given the measured covariates, \(C \perp\!\!\!\perp (T,J) \mid {X}\), which reduces to \(C \perp\!\!\!\perp T \mid {X}\) when $M=1$, (2) the positivity condition holds, so that \(G(\tau\mid\mathbf{x})\geq\epsilon>0\) for all \(\mathbf{x}\) in the support of the baseline covariates, and (3) the censoring model is correctly specified, meaning \(\widehat{G}(u\mid\mathbf{x})\) is a consistent estimator of \(G(u\mid\mathbf{x})\).
Under these assumptions, we define the proposed CIPCW-based estimators of time-dependent prevalence, sensitivity and specificity of cause $k$ as\cite{blanche2013estimating, lunceford2004stratification}:

\begin{align}
\widehat{\mathrm{Prev}}_{\tau}^{(k)}
&=
\frac{
\sum_{i=1}^{n}
I(\widetilde{T}_i\leq\tau,\Delta_i=k) W_i
}{
\sum_{i=1}^{n}
\left[
I(\widetilde{T}_i\leq\tau,\Delta_i\neq 0) W_i
+
I(\widetilde{T}_i>\tau) W_i^\prime
\right]
} \notag
\\[1.5ex]
\widehat{\mathrm{Se}}_{\tau}^{(k)}
&=
\frac{
\sum_{i=1}^{n}
I(\pi_i^{(k)}>z,\widetilde{T}_i\leq\tau,\Delta_i=k) W_i
}{
\sum_{i=1}^{n}
I(\widetilde{T}_i\leq\tau,\Delta_i=k) W_i
} \label{eq:cipcw}
\\[1.5ex]
\widehat{\mathrm{Sp}}_{\tau}^{(k)}
&=
\frac{
\sum_{i=1}^{n}
I(\pi_i^{(k)}\leq z)
\left[
I(\widetilde{T}_i>\tau) W_i^\prime
+
I(\widetilde{T}_i\leq\tau,\Delta_i\notin\{0,k\}) W_i
\right]
}{
\sum_{i=1}^{n}
\left[
I(\widetilde{T}_i>\tau) W_i^\prime
+
I(\widetilde{T}_i\leq\tau,\Delta_i\notin\{0,k\}) W_i
\right]
} \notag
\end{align}
When $M = 1$, $\Delta_i = k$ reduces to $\Delta_i = 1$, which indicates an observed event, and $\widehat{\mathrm{Sp}}_{\tau}^{(k)}$ is simplified by removing the term \(I(\widetilde{T}_i\leq\tau,\Delta_i\notin\{0,k\})\,W_i\), yielding the CIPCW-based estimators for a standard survival outcome. We show in Supporting Information, Section S6, that these estimators are consistent and asymptotically unbiased under Assumptions (1)--(3) stated above. Consequently, substituting the estimators in Equation~\eqref{eq:cipcw} into Equation~\eqref{eq:survival_nb} yields consistent and asymptotically unbiased estimators of survival NB for both the standard survival and competing-risks settings.

Confidence intervals (CIs) for the proposed NB estimators are obtained using bootstrap resampling~\cite{blanche2013estimating}, since deriving closed-form variance expressions is complex~\cite{martinussen2006dynamic}. CIs are used only to assess empirical coverage in the simulation study and not as a metric for quantifying uncertainty in NB, because CIs around NB do not naturally translate into clinical decision-making.\cite{Vickers2023DecisionCurve}

\subsection{Software Implementation}
We implemented the proposed estimator in a fork of the \texttt{dcurves} R package\cite{sjoberg2025dcurves}, with accompanying vignettes, available at \url{https://github.com/amirfarahmand0/dcurves/tree/feature/cipcw}. Our implementation supports two ways of obtaining the censoring weights: the conditional censoring survival probabilities can either be estimated internally using a Cox proportional hazards model (default) or supplied externally from a user-specified censoring model.

\section{Simulation Study}

We conducted Monte Carlo simulation studies to evaluate the proposed CIPCW-based approach for NB-based external validation of survival prediction models in both the standard survival and competing-risk settings. Since the conventional approach is valid only under non-informative censoring and the CIPCW-based approach relies on a correctly specified censoring model, we considered scenarios with different censoring mechanisms, censoring rates, and censoring model specifications. We describe the simulation studies below using the ADEMP framework (aims, data-generating mechanisms, estimands, methods, and performance measures).\cite{morris2019} Full details, including the exact specification of the data-generating mechanisms and the rationale for the number of replications and the validation sample size, are provided in Supporting Information, Section S1.

\subsection{Aims}
The primary aim of the simulation study was to compare the bias, precision, and nominal CI coverage of the proposed CIPCW-based approach with those of the conventional approach under non-informative and informative censoring. A secondary aim was to assess the sensitivity of the CIPCW-based approach to misspecification of the censoring model.

\subsection{Data-generating mechanisms}
Event times were generated from exponential proportional hazards models based on two correlated standard normal covariates, $Z_1$ and $Z_2$ (correlation=0.5). In the standard survival setting, the event hazard was $\lambda_T\exp(Z_1+Z_2)$, with $\lambda_T$ calibrated at a 5-year time horizon to yield a fixed 30\% event rate, $P(T\leq 5)=0.30$. In the competing-risk setting, a second latent event time for the competing event was introduced alongside the primary event of interest, with cause-specific hazards $\lambda_1\exp(Z_1+Z_2)$ and $\lambda_2\exp(-0.5Z_1+0.75Z_2)$ jointly calibrated to target a 30\% cumulative incidence for the primary event and 10\% for the competing event at the same horizon.

Censoring times were generated from an exponential model with hazard $\lambda_C\exp(\mathrm{LP}_C)$ under two censoring settings: (1) non-informative censoring, $\mathrm{LP}_C=0$, and (2) informative censoring with a linear predictor, $\mathrm{LP}_C=\log(2.65)(Z_1+Z_2)$, in which the censoring hazard increased with both $Z_1$ and $Z_2$ (hazard ratio 2.65 per unit increase in each). In a sensitivity analysis, we also considered informative censoring with a nonlinear predictor, $\mathrm{LP}_C=1.5\tanh(Z_1)-2\exp(-Z_2^2)$. In all settings, $\lambda_C$ was calibrated to target censoring rates of 10\%, 20\%, and 30\%, defined as the proportion of individuals censored before experiencing an event and before the 5-year horizon. These mechanisms are modeling choices of the simulation, whereas the CIPCW-based approach itself only requires Assumptions (1)--(3) (Section~2.3). For each scenario and censoring rate, a reference population of one million individuals was generated, from which we drew 1,000 validation datasets without replacement, each sized to yield 100 expected primary events observed before the 5-year horizon, so that each validation dataset had enough events. The number of replications was chosen to keep the Monte Carlo standard error of the bias below 0.005.\cite{morris2019}

\subsection{Prediction model}
A fixed survival prediction model was developed on an independent training set of 1,000 individuals generated under non-informative censoring. In the standard survival setting, this was a Cox proportional hazards model including $Z_1$ and $Z_2$, and in the competing-risk setting, cause-specific Cox models for the primary and competing events were combined to predict the cumulative incidence of the primary event.

\subsection{Estimand}
The estimand was the NB of the decision strategy based on the prediction model at the 5-year horizon, evaluated at decision thresholds of 0.01, 0.10, 0.25, 0.50, and 0.75. For each scenario and censoring rate, the \textit{true} NB of this prediction model was computed in the reference population from the event times before censoring was applied.

\subsection{Methods}
In each validation dataset, we compared the conventional KM-based approach for the standard setting and AJ-based approach for the competing-risk setting against the proposed CIPCW approach. To assess sensitivity to censoring model specification, censoring weights were estimated using two variations, a correctly specified Cox model including $Z_1$ and $Z_2$, and a misspecified Cox model including $Z_1$ only. In the sensitivity analysis, because the linear Cox censoring model is misspecified under nonlinear informative censoring, censoring weights were additionally estimated using a Cox generalized additive model (Cox GAM) with penalized cubic regression splines for $Z_1$ and $Z_2$, which can capture nonlinear covariate effects.\cite{wood2016} The Cox GAM was also applied under non-informative and linear informative censoring to assess its performance when this flexibility is not needed. For all censoring models, estimated conditional censoring survival probabilities were truncated at 0.02 to avoid extreme weights. To assess coverage, which is the standard check that the bootstrap captures the variability of the estimators, we constructed 90\% CIs from the 5th and 95th percentiles of 500 nonparametric bootstrap resamples of each validation dataset, and the censoring model was refitted in every resample, whereas the prediction model was kept fixed. 

\subsection{Performance measures}
Performance was summarized by the relative bias, empirical standard error, and relative root mean squared error (RMSE), with the mean bias and RMSE across 1,000 replications expressed as a percentage of the true NB,\cite{morris2019} together with the empirical coverage of the 90\% bootstrap CIs compared with the nominal 90\% level. Supporting Information, Tables S2--S5 report the relative and absolute bias and RMSE, empirical standard errors, coverage, and their Monte Carlo standard errors.

\subsection{Results}
\begin{figure}[!htbp]
    \centering
    \includegraphics[width=\textwidth,height=0.82\textheight,keepaspectratio]{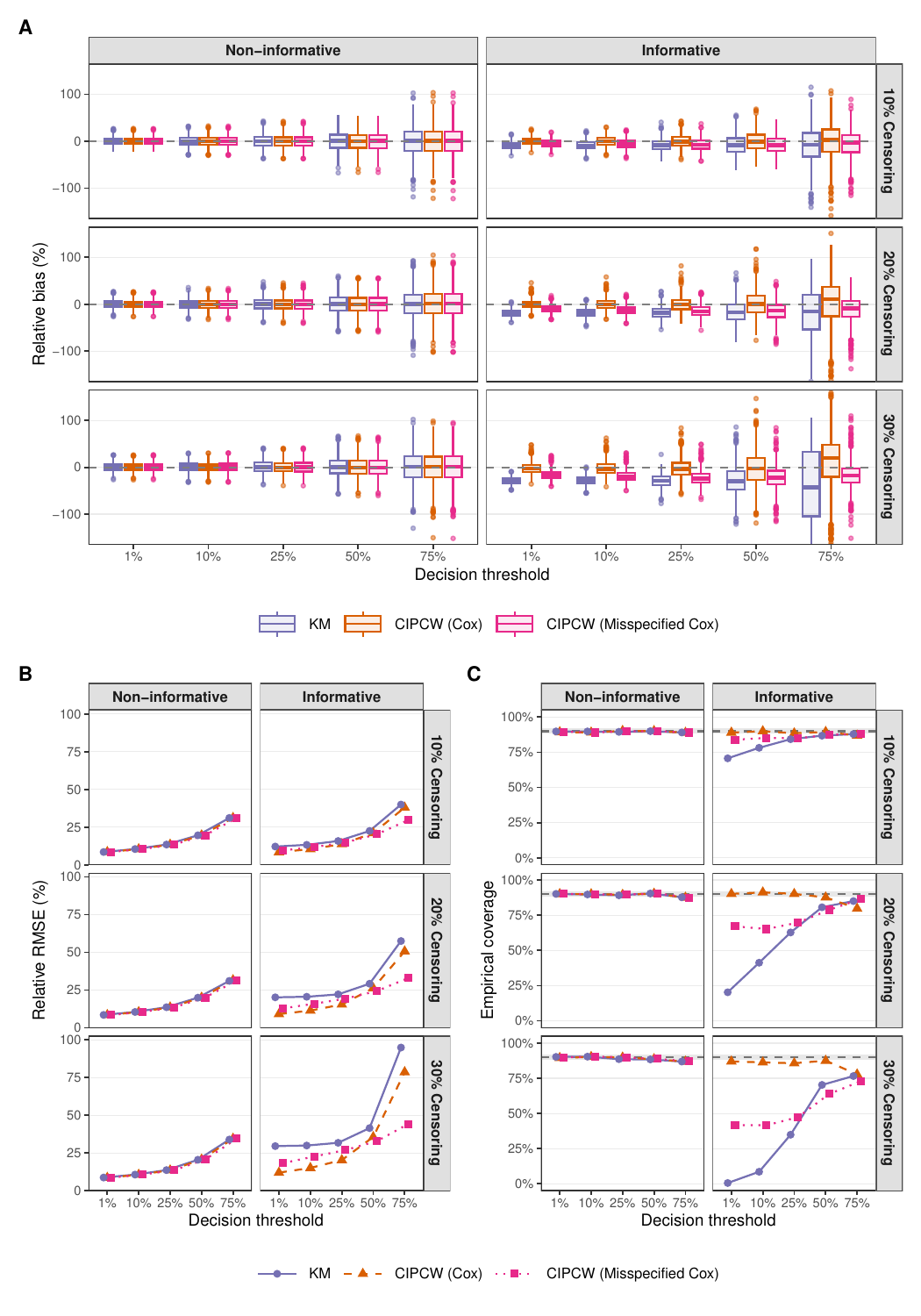}
    \caption{Simulation results for the standard survival setting ($M=1$) at 10\%, 20\%, and 30\% censoring, based on 1,000 Monte Carlo replications. (A) Distribution of the relative bias in the estimated NB (estimated minus true NB, as a percentage of the true NB) across decision thresholds. (B) Relative root mean squared error (RMSE) of the estimated NB. (C) Empirical coverage of the 90\% bootstrap confidence intervals. The dashed line marks the nominal level and the shaded band its Monte Carlo interval. Informative censoring refers to censoring with a linear predictor. \textit{Abbreviations:} CIPCW = conditional inverse probability of censoring weighting, KM = Kaplan--Meier, NB = net benefit.}
    \label{fig:main_standard}
\end{figure}

\begin{figure}[!htbp]
    \centering
    \includegraphics[width=\textwidth,height=0.82\textheight,keepaspectratio]{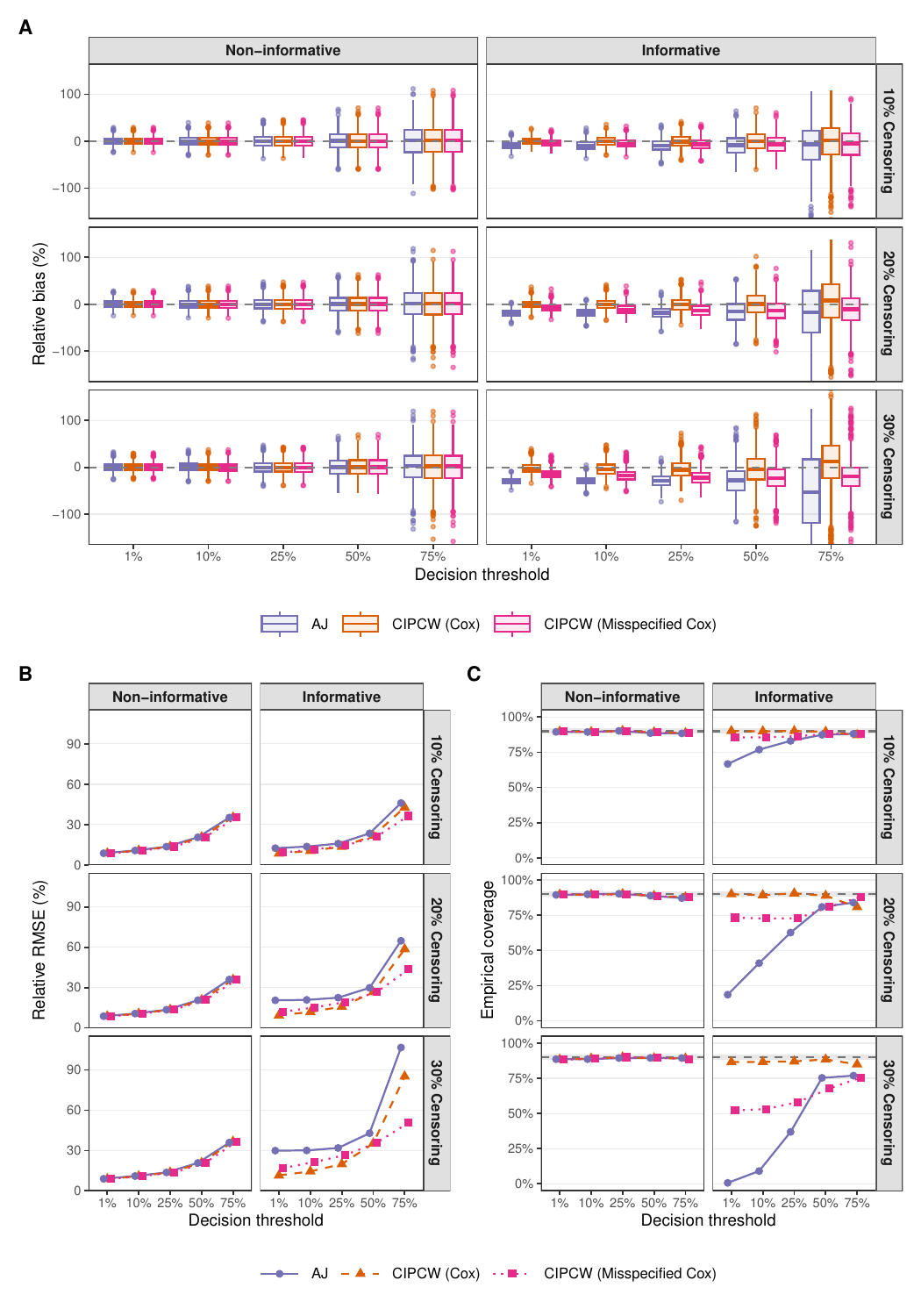}
    \caption{Simulation results for the competing-risk setting ($M=2$) at 10\%, 20\%, and 30\% censoring, based on 1,000 Monte Carlo replications. (A) Distribution of the relative bias in the estimated NB (estimated minus true NB, as a percentage of the true NB) across decision thresholds. (B) Relative root mean squared error (RMSE) of the estimated NB. (C) Empirical coverage of the 90\% bootstrap confidence intervals. The dashed line marks the nominal level and the shaded band its Monte Carlo interval. Informative censoring refers to censoring with a linear predictor. \textit{Abbreviations:} AJ = Aalen--Johansen, CIPCW = conditional inverse probability of censoring weighting, NB = net benefit.}
    \label{fig:main_competing}
\end{figure}

Under non-informative censoring, the CIPCW-based approach performed similarly to the conventional approaches in both the standard and competing-risk settings, with negligible bias for all approaches across thresholds and censoring rates (below 1\% of the true NB, Panel A of Figures~\ref{fig:main_standard} and \ref{fig:main_competing}). Estimating the censoring weights did not increase RMSE or reduce precision, as the relative RMSE (Panel B) and empirical standard errors (Supporting Information, Tables S2 and S3) of the CIPCW-based estimators were nearly identical to those of the conventional estimators. The 90\% bootstrap CIs of all approaches achieved coverage close to the nominal level (approximately 89\% on average, Panel C of Figures~\ref{fig:main_standard} and \ref{fig:main_competing}).

In contrast, when censoring was informative, the correctly specified CIPCW-based estimator remained largely unbiased (within 4\% of the true NB), whereas the KM-based and AJ-based estimators showed substantial bias that increased with the censoring rate, underestimating the true NB by up to 29\% at thresholds up to 0.50, and substantially more at the 0.75 threshold, where the true NB was smallest (Panel A of Figures~\ref{fig:main_standard} and \ref{fig:main_competing}, Supporting Information, Tables S2 and S3). The relative RMSE of the CIPCW-based estimator was lower at every threshold and censoring rate (Panel B of Figures~\ref{fig:main_standard} and \ref{fig:main_competing}), although it generally had larger empirical standard errors than the KM-based and AJ-based estimators (Supporting Information, Tables S2 and S3), as the bias of the conventional approaches outweighed their smaller empirical standard errors. Consequently, the KM-based and AJ-based approaches showed considerable undercoverage (below 1\% at the lowest threshold with 30\% censoring), while the correctly specified CIPCW-based approach generally maintained coverage close to 90\% (at least 85\% at thresholds up to 0.50, Panel C of Figures~\ref{fig:main_standard} and \ref{fig:main_competing}). The exception was the highest threshold at higher censoring rates, where the few individuals classified as high risk were also those most likely to be censored, so their weights were large and often truncated (Supporting Information, Table S7).

When the censoring model was misspecified, the CIPCW-based estimator showed noticeable bias under informative censoring (up to 23\% of the true NB), although its bias remained smaller than that of the KM-based and AJ-based estimators (Panel A of Figures~\ref{fig:main_standard} and \ref{fig:main_competing}). Its relative RMSE was lower than that of the conventional estimators at all thresholds, and at the highest threshold it was even lower than that of the correctly specified CIPCW-based estimator, because omitting $Z_2$ resulted in smaller weights for these high-risk individuals, which reduced variability at the cost of bias (Panel B of Figures~\ref{fig:main_standard} and \ref{fig:main_competing}). Misspecification of the censoring model led to varying degrees of undercoverage (as low as 42\%), which was most visible at low decision thresholds, where NB was more sensitive to errors in the estimation of prevalence, sensitivity, and specificity because nearly all individuals were classified as high risk (Panel C of Figures~\ref{fig:main_standard} and \ref{fig:main_competing}).

In the sensitivity analysis, under nonlinear informative censoring, the CIPCW-based estimator with a linear Cox censoring model remained biased, although mostly less than the KM-based and AJ-based estimators, and its coverage fell below the nominal level at low thresholds. Estimating the censoring weights with the Cox GAM substantially reduced this bias (within 7\% of the true NB at thresholds up to 0.50, compared with up to 18\% for the KM-based and AJ-based estimators) and maintained coverage close to 90\% at thresholds up to 0.50, although it showed some positive bias and undercoverage at the 0.75 threshold. Under non-informative and linear informative censoring, the Cox GAM performed similarly to the correctly specified Cox model, indicating that its additional flexibility came at little cost in precision (Supporting Information, Section S3, Figures S1 and S2, and Tables S4 and S5).

\section{Case Study}
\subsection{Prediction model} 
To illustrate the proposed CIPCW-based approach in practice, we externally evaluated the clinical utility of a recurrence-free survival (RFS) prediction model previously developed by Royston and Altman in the Rotterdam cohort\cite{royston2013external}. RFS was defined as the time from primary surgery to the earlier of disease recurrence or death from any cause. The model included age modeled using a fractional polynomial transformation, menopausal status, tumor size, number of positive lymph nodes, estrogen receptor level, and hormonal treatment status. Further details on the model specification can be found in the study by Royston and Altman~\cite{royston2013external}, and the model can be reproduced using the \texttt{rotterdam} dataset in the R \texttt{survival} package. The code for the case study is available at \url{https://github.com/amirfarahmand0/paper-RobustDCA}.

\subsection{Decision problem and validation cohort}
We considered the decision problem of whether adopting this model to guide the use of an intensified adjuvant systemic therapy would do more good than harm in German women with node-positive breast cancer. To address this, we applied the model to the German Breast Cancer Study Group (GBSG) cohort\cite{schumacher1994randomized}, an independent validation sample taken to be representative of this population, and obtained each patient's predicted 5-year risk of recurrence or death. Using DCA, we compared a strategy in which the therapy is offered to patients whose predicted risk exceeds a given threshold against the default strategy of treating all patients~\cite{vangeloven2022validation}. We evaluated decision thresholds between \(14\%\) and \(23\%\), consistent with clinically relevant thresholds previously considered for chemotherapy decisions based on the 5-year risk of breast cancer recurrence or death\cite{mclernon2023assessing}.

The GBSG trial enrolled 720 patients with primary node-positive breast cancer between 1984 and 1989, of whom all 686 with complete information on age, menopausal status, tumor size, number of positive lymph nodes, estrogen receptor level, and hormonal treatment status comprised the validation cohort\cite{schumacher1994randomized}, as the model was developed in the Rotterdam cohort. Over a maximum follow-up of 7.3 years, 299 (43.6\%) individuals experienced an RFS event (recurrence or death) and 387 (56.4\%) were censored and by the 5-year prediction horizon, 285 (41.5\%) had experienced an RFS event and 280 (40.8\%) had been censored.

\subsection{Censoring mechanism and censoring model}

\begin{figure}[!htbp]
    \centering
    \includegraphics[width=0.8\textwidth]{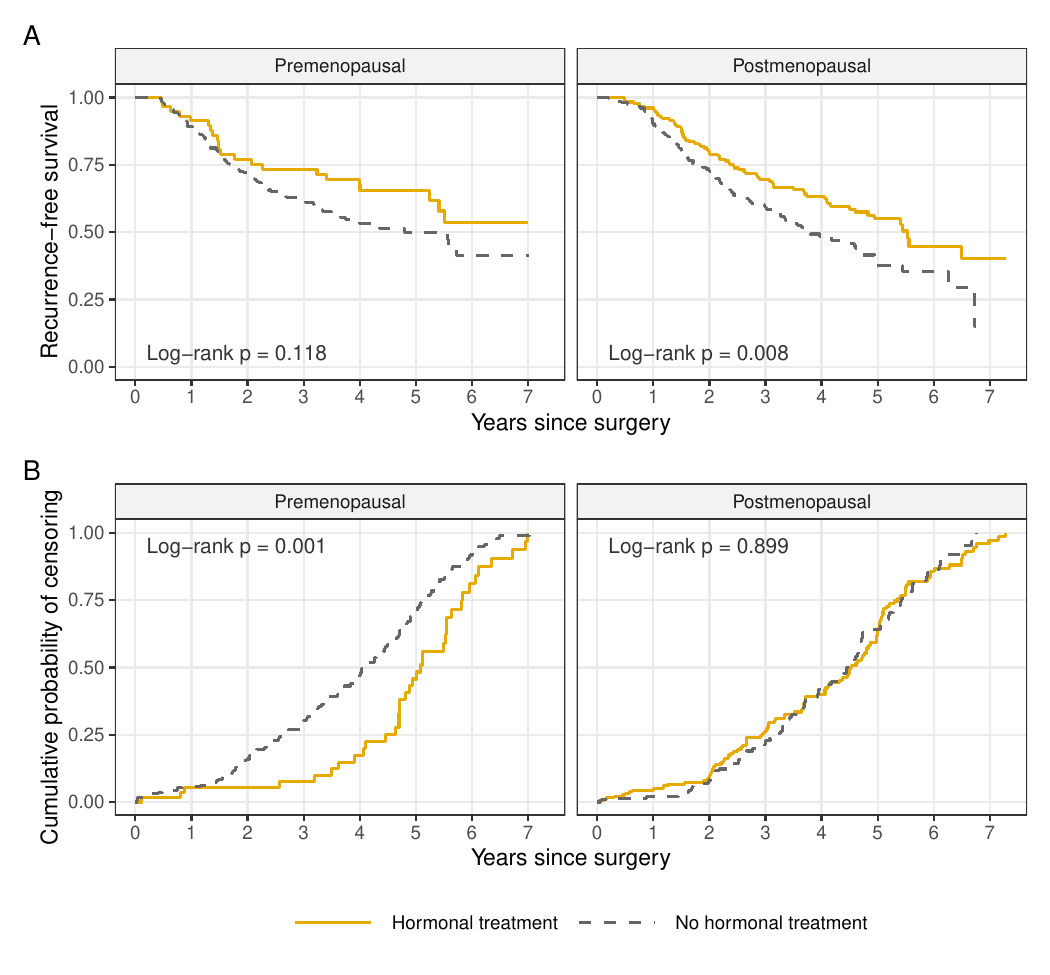}
    \caption{Recurrence-free survival (A) and cumulative probability of censoring (B) according to hormonal treatment status, shown separately for premenopausal and postmenopausal women in the GBSG validation cohort. Both are estimated using the Kaplan--Meier estimator, applied to the censoring times in panel B. $p$-values are from log-rank tests comparing women with and without hormonal treatment within each menopausal group. \textit{Abbreviation:} GBSG = German Breast Cancer Study Group.}
    \label{fig:gbsg_censoring}
\end{figure}

Recurrence-free survival differed by hormonal treatment status in the whole cohort ($p=0.003$), with a similar pattern in both menopausal groups (Panel A of Figure~\ref{fig:gbsg_censoring}). In the GBSG trial, hormonal treatment consisted of tamoxifen for two years,\cite{schumacher1994randomized} and premenopausal women were randomized to tamoxifen only until December 1986.\cite{colleoni2002duration} As a result, premenopausal women enrolled after this date did not receive hormonal treatment, and because follow-up ended at the same time for all patients, they had shorter potential follow-up than premenopausal women with hormonal treatment. As such, censoring differed by hormonal treatment status among premenopausal women ($p=0.001$) but not among postmenopausal women, who were randomized to tamoxifen throughout the trial ($p=0.90$, Panel B of Figure~\ref{fig:gbsg_censoring}). Since hormonal treatment was associated with both recurrence-free survival and censoring, censored individuals were unlikely to be representative of those remaining under follow-up, corroborating the violation of the non-informative censoring assumption underlying the KM-based approach.

To account for this informative censoring, we estimated the CIPCW weights using a Cox proportional hazards model for the censoring times that included baseline covariates recorded in the trial (age, menopausal status, tumor size, number of positive lymph nodes, estrogen receptor level, and hormonal treatment status) and an interaction between hormonal treatment and menopausal status, allowing the association between hormonal treatment and censoring to differ by menopausal status as implied by the trial design. In this model, and consistent with the trial design, hormonal treatment was associated with a lower censoring hazard among premenopausal women (hazard ratio 0.53, 95\% CI: 0.36--0.76) but not among postmenopausal women (hazard ratio 0.99, 95\% CI: 0.76--1.30). The global Schoenfeld residual test did not indicate a violation of the proportional hazards assumption, and the weights stayed within a reasonable range (median 1.25, maximum 5.04). The full censoring model, including all hazard ratios and the proportional hazards tests, is presented in Supporting Information, Section S4 and Table S6.

\begin{figure}[!htbp]
    \centering
    \includegraphics[width=\textwidth]{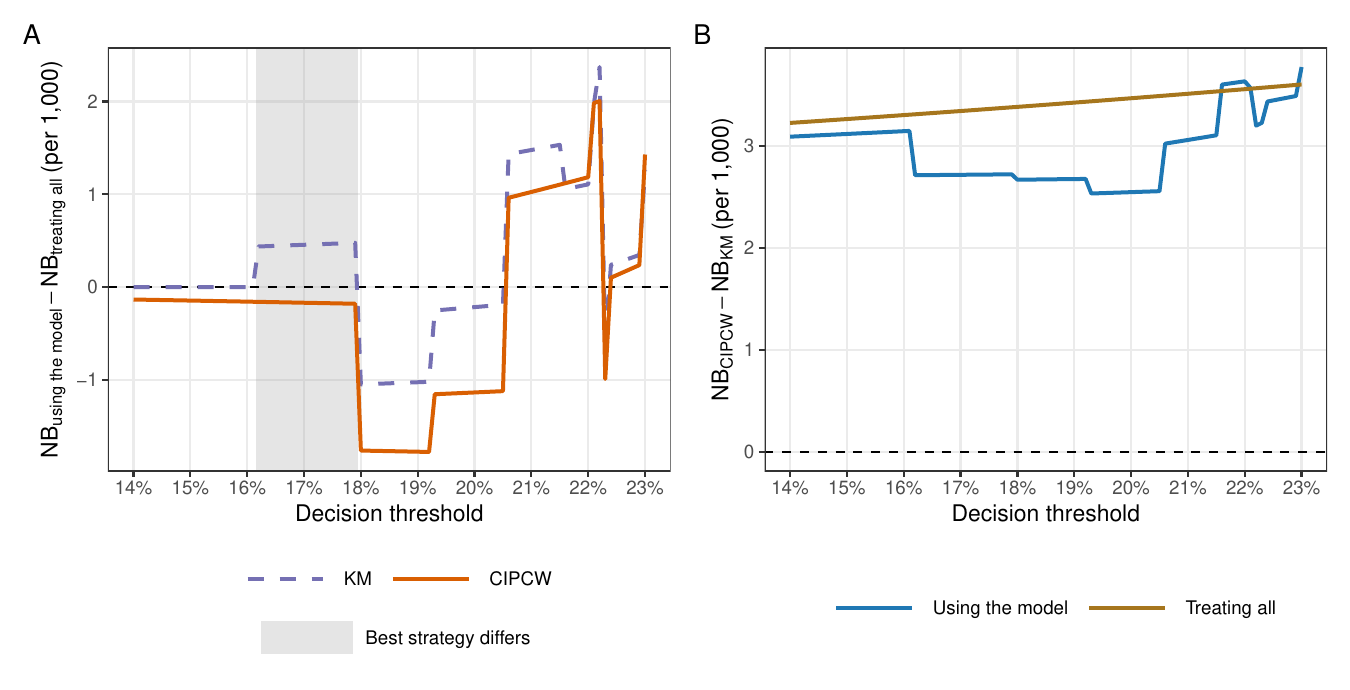}
    \caption{Impact of accounting for informative censoring on clinical decision-making in the GBSG validation cohort. (A) Difference in NB between using the model and treating all patients per 1,000 individuals, estimated using the KM-based and CIPCW-based approaches. Values above zero favor using the model, whereas values below zero favor treating all patients. The shaded region indicates decision thresholds at which the best strategy differed between the two approaches. (B) Difference in estimated NB between the CIPCW-based and KM-based approaches per 1,000 individuals, shown separately for using the model and treating all patients. Positive values indicate a higher estimated NB when informative censoring is accounted for using CIPCW. \textit{Abbreviations:} CIPCW = conditional inverse probability of censoring weighting, GBSG = German Breast Cancer Study Group, KM = Kaplan--Meier, NB = net benefit.}
    \label{fig:gbsg_decision_impact}
\end{figure}

\subsection{Case study results}      

Using the conventional KM-based approach, the best strategy varied across the prespecified threshold range (Panel A of Figure~\ref{fig:gbsg_decision_impact}). Using the model was best at thresholds between 16.2\% and 17.9\% and above 20.6\%, whereas treating all patients was best between 18.0\% and 20.5\%. Below 16.2\%, the two strategies had almost identical NB. However, these estimates rely on the assumption that censoring is non-informative, which is questionable in this cohort given how the trial was conducted.

Accounting for informative censoring via the CIPCW estimator changed the best strategy at thresholds between 16.2\% and 17.9\% (shaded region in Panel A of Figure~\ref{fig:gbsg_decision_impact}), where the KM-based approach favored using the model whereas the CIPCW-based approach favored treating all patients. Outside this range, both approaches led to the same decision. Across the threshold range, the CIPCW-based NB estimates were higher than the KM-based estimates by 2.5 to 3.8 per 1,000 individuals for using the model and by 3.2 to 3.6 per 1,000 for treating all patients (Panel B of Figure~\ref{fig:gbsg_decision_impact}), a difference equivalent to 2.5 to 3.8 women per 1,000 who would experience recurrence or death within 5 years being appropriately treated, without any increase in unnecessary treatment. Thus, for decision thresholds in this range, ignoring informative censoring can lead to using the model to guide intensified therapy, whereas accounting for it showed that treating all women was best.

\section{Discussion}

In this study, we introduced a robust NB estimator leveraging the CIPCW methodology to correct bias in clinical utility estimation when follow-up depends on patient characteristics associated with the outcome. Our simulation studies showed that the conventional approaches can produce substantially biased NB estimates, especially at higher censoring rates, which in our simulations made the prediction model appear less useful than it truly was. In contrast, the proposed CIPCW-based estimator largely corrected this bias, with a lower RMSE and coverage close to the nominal level, while performing similarly to the conventional approaches when censoring was non-informative. However, when the censoring model was misspecified, its bias and undercoverage increased, although it remained less biased than the conventional approaches. In the GBSG case study, accounting for potentially informative censoring changed the best strategy over part of the clinically relevant threshold range, demonstrating that conventional DCA for survival outcomes may lead to suboptimal clinical decisions.

This methodological advancement is consistent with the broader development of performance metrics for survival prediction models, where accounting for informative censoring has become increasingly recognized.~\cite{prince2025estimation, lillelund2025overcoming} Similar to our findings, several studies have shown that failing to account for potential informative censoring can lead to misleading evaluation of survival prediction models.~\cite{lillelund2024stop, kvamme2023brier} Prediction error measures based on the Brier score provide an early example, with Graf et al.~\cite{Graf1999} demonstrating the impact of censoring on prediction error estimation and Gerds and Schumacher~\cite{gerds2006consistent} establishing consistent CIPCW-based estimation of the Brier score under covariate-dependent censoring. Similar extensions have been developed for discrimination measures, including time-dependent sensitivity and specificity~\cite{blanche2013estimating} and the C-index.~\cite{gerds2013estimating} However,  best practice guidelines recommend only reporting how censoring is handled when assessing NB.~\cite{collins2024tripodai, wolff2019probast} Similarly, the STRATOS guidance for survival models estimates NB under non-informative censoring, despite noting that loss to follow-up may depend on patient characteristics.~\cite{mclernon2023assessing} Future guidelines should recommend evaluating the censoring mechanism and reporting censoring-adjusted NB estimates, at least as a sensitivity analysis. This requires event-history data and an additional censoring model that should itself be validated.

A primary strength of the proposed CIPCW framework is its flexibility in modeling the censoring mechanism, using approaches ranging from Cox regression to machine learning methods,~\cite{prince2025estimation} and our \texttt{dcurves} implementation accepts censoring probabilities from any user specified model. However, data-driven approaches should be used with caution, as weights estimated by flexible machine learning methods can be imprecise because these methods need large samples to estimate censoring probabilities accurately, and these errors pass directly into the weighted estimate, leading to bias and poor confidence interval coverage.~\cite{naimi2021challenges} Similarly, Prince et al.~\cite{prince2025estimation} found that poorly tuned machine learning censoring models can substantially bias IPCW-based performance estimates, and that under covariate-dependent censoring, the censoring model is best estimated in the validation data, as in our framework. More generally, we suggest specifying the censoring model before data analysis in collaboration with subject matter experts, based on why individuals are censored. We recommend the causal DAG approach described by Hernán et al.~\cite{hernan2004structural} for this purpose. Such a DAG helps identify variables associated with both censoring and event time, which should be included in the censoring model, as omitting them can leave residual selection bias.~\cite{hernan2004structural, xuan2024safer}

Despite its advantages, the CIPCW-based approach has some limitations, and caution is needed when applying it. First, it relies on the assumption that censoring is independent of the event time given the measured covariates, which cannot be verified from the observed data, and it cannot correct bias from unmeasured factors related to both censoring and the event.~\cite{lillelund2025overcoming} Covariate selection for the censoring model should be based on causal knowledge.~\cite{hernan2004structural} Second, the CIPCW-based approach provides no theoretical protection against misspecification of the censoring model itself, even though in our simulations the estimates seemed relatively robust to such misspecification. A censoring model that captures only part of the censoring mechanism, for example through covariates correlated with the omitted drivers of censoring, may still reduce bias, but this is not guaranteed. Third, our simulation study used exponential event and censoring times with constant hazards, and the case study used a single cohort, so the performance of the approach under time-varying hazards and in other settings remains to be assessed. Finally, as with other inverse probability weighting methods, near violations of the positivity assumption, where some covariate groups have very low probabilities of remaining uncensored, can produce extreme weights that increase variance and lead to unstable estimates, as observed at the highest threshold in our simulation.~\cite{howe2011limitation}

To address these limitations, subsequent work should explore doubly robust estimation techniques~\cite{wen2021multiply, bang2005doubly} to protect against censoring model misspecification. Copula-based methods~\cite{lillelund2024stop, lillelund2025overcoming} could address complex dependence structures between censoring and event times, and the framework could be extended to censoring weights based on time-dependent covariates.~\cite{robins2000correcting} Regarding uncertainty, we intentionally did not report confidence intervals in the case study, as the use of standard inferential methods to express the uncertainty of a decision-theoretic measure such as NB is questionable, and whether and how to quantify uncertainty around DCA remains controversial.~\cite{kerr2019importance, capogrosso2019systematic} Instead, future studies could integrate the proposed framework with value of information analysis, which expresses this uncertainty in terms of its consequences for decisions.~\cite{evpival, evsival}

Informative censoring can make a survival prediction model look more useful than it is, or less. If the bias is ignored, clinicians risk making the wrong call about whether to act on the model's predictions. The proposed CIPCW-based NB estimator corrects this bias by accounting for why patients are censored, and it is freely available in a fork of the \texttt{dcurves} R package. When censoring may be informative, we recommend its routine use in decision curve analysis for survival outcomes.

\section*{Code/Data Availability}

All the analysis code and data for this study, including the simulation and case study, are available from https://github.com/amirfarahmand0/paper-RobustDCA.

\section*{Funding/Support}
This study was supported by the Banting Postdoctoral Fellowship from the Canadian Institutes of Health Research (Lee).

\section*{Role of the Funder/Sponsor}
The content is solely the responsibility of the authors and does not necessarily represent the official views of the Canadian Institutes of Health Research. The funding organizations had no role in the design and conduct of the study, collection, management, analysis, and interpretation of the data, preparation, review, or approval of the manuscript, and decision to submit the manuscript for publication.

\section*{Author Contributions}
TYL conceived the study question. AF, AS, and TYL designed the study and developed the statistical methodology. AF collected the data, implemented the software, and performed the formal data analysis. AF drafted the manuscript. AS and TYL supervised the project. All authors reviewed, revised, and approved the final manuscript.

\section*{Conflicts of Interest}
The authors declare no conflicts of interest.

\section*{Use of AI Tools}
Claude (Anthropic) was used to assist with language refinement and consistency in writing style across the manuscript, and with implementing parts of the analysis code. The study design, methodology, and analytical decisions are the authors' own, and the authors retain full responsibility for the scientific content, interpretation, and conclusions of the manuscript.

\bibliographystyle{SageV}
\bibliography{references}

\clearpage
\newgeometry{left=2cm,right=2cm,top=1.8cm,bottom=1.8cm}
\singlespacing
\setcounter{section}{0}\setcounter{figure}{0}\setcounter{table}{0}\setcounter{equation}{0}
\renewcommand{\thesection}{S\arabic{section}}
\renewcommand{\thefigure}{S\arabic{figure}}
\renewcommand{\thetable}{S\arabic{table}}
\renewcommand{\theequation}{S\arabic{equation}}
\captionsetup{labelfont=bf,labelsep=period,font=small}
\setlength{\LTcapwidth}{\textwidth}
\setlength{\parindent}{0pt}

\begin{center}
{\Large\bfseries Supporting Information}\\[0.3em]
{\small Robust Decision Curve Analysis for Survival Risk Prediction Models in the Presence of Informative Censoring}
\end{center}

\section{Simulation design}\label{sec:design}

\begin{table}[H]
\centering\footnotesize
\caption{Simulation design. (A) Data-generating mechanisms, prediction model, validation datasets and bootstrap. (B) Specification of the censoring models used to estimate the CIPCW weights under each censoring mechanism. \textit{Abbreviations:} CIPCW = conditional inverse probability of censoring weighting, GAM = generalized additive model, NB = net benefit.}\label{tab:design}
\renewcommand{\arraystretch}{1.08}
\begin{tabularx}{\textwidth}{@{}>{\raggedright\arraybackslash}p{3.4cm} Y@{}}
\toprule
\multicolumn{2}{@{}l}{\textbf{(A) Data-generating mechanisms}}\\
\midrule
Covariates & $(Z_1, Z_2)^\top \sim \mathcal{N}_2(\mathbf{0}, \Sigma)$ with unit variances and correlation 0.5 \\
Event time, standard survival & $T \sim \mathrm{Exp}\{\lambda_T \exp(Z_1 + Z_2)\}$, with $\lambda_T$ calibrated so that $P(T \le 5) = 0.30$ \\
Event times, competing risks & $T_1 \sim \mathrm{Exp}\{\lambda_1 \exp(Z_1 + Z_2)\}$ and $T_2 \sim \mathrm{Exp}\{\lambda_2 \exp(-0.5 Z_1 + 0.75 Z_2)\}$, $T = \min(T_1, T_2)$, with $J = 1$ if $T_1 < T_2$ and $J = 2$ otherwise. $\lambda_1$ and $\lambda_2$ calibrated jointly so that $P(T \le 5, J = 1) = 0.30$ and $P(T \le 5, J = 2) = 0.10$ \\
Censoring time & $C \sim \mathrm{Exp}\{\lambda_C \exp(\mathrm{LP}_C)\}$, with $\lambda_C$ calibrated so that $P\{C < \min(T, 5)\} = 0.10$, $0.20$ or $0.30$ \\
\quad Non-informative & $\mathrm{LP}_C = 0$ \\
\quad Linear informative & $\mathrm{LP}_C = \log(2.65)\, Z_1 + \log(2.65)\, Z_2$ \\
\quad Nonlinear informative & $\mathrm{LP}_C = 1.5 \tanh(Z_1) - 2 \exp(-Z_2^2)$ (sensitivity analysis) \\
Prediction model & Fitted on a training set of 1,000 individuals with non-informative censoring: a Cox model with $Z_1$ and $Z_2$ (standard survival), or cause-specific Cox models with $Z_1$ and $Z_2$ combined to predict the cumulative incidence of cause 1 (competing risks). Predicted risk at $\tau = 5$ \\
Reference population & $10^6$ individuals per scenario and censoring rate. The true NB was computed from the event times before censoring \\
Replications & 1,000 validation datasets per scenario and censoring rate, drawn without replacement from the reference population, to keep the Monte Carlo standard error of the bias below 0.005 \\
Validation sample size & $n = \lceil 100/p \rceil$, where $p$ is the proportion of the reference population with an observed primary event by $\tau = 5$. This gives about 100 observed primary events in each validation dataset regardless of the censoring rate, so that the CIPCW-based approach has enough events and datasets with very few events are avoided \\
Confidence intervals & 5th and 95th percentiles of 500 nonparametric bootstrap resamples, enough for a stable estimation of these percentiles. The censoring model was refitted in each resample and the prediction model was kept fixed \\
\midrule
\end{tabularx}

\vspace{0.2em}
\begin{tabularx}{\textwidth}{@{}>{\raggedright\arraybackslash}p{3.4cm} Y Y Y@{}}
\multicolumn{4}{@{}l}{\textbf{(B) Censoring models}}\\
\midrule
Censoring model & Non-informative & Linear informative & Nonlinear informative \\
\midrule
Cox, $Z_1 + Z_2$ (linear effects) & Not applicable$^a$ & Correct & Misspecified (functional form) \\
Cox, $Z_1$ only & Not applicable$^a$ & Misspecified (omitted covariate) & Misspecified (omitted covariate and functional form) \\
GAM, $s(Z_1) + s(Z_2)$ & Not applicable$^a$ & Correct & Correct up to smoothing$^b$ \\
\bottomrule
\end{tabularx}

\vspace{0.3em}
\parbox{\textwidth}{\footnotesize $^a$Censoring does not depend on $Z_1$ or $Z_2$, so no covariate-dependent censoring model is needed. $^b$The nonlinear mechanism is additive in $Z_1$ and $Z_2$, so the GAM can approximate it up to the smoothing. The GAM was fitted with penalized cubic regression splines (basis dimension 4) using the Cox proportional hazards family of the \texttt{mgcv} R package. For all censoring models, the estimated conditional censoring survival probabilities were truncated at 0.02.}
\end{table}

\clearpage
\section{Simulation results: numerical summaries}\label{sec:main}

\FloatBarrier
{\scriptsize\setlength{\tabcolsep}{2.2pt}
\begin{longtable}{@{}r r l r r r r r r@{}}
\caption{Performance of the net benefit estimators in the standard survival setting at 10\%, 20\% and 30\% censoring, based on 1,000 Monte Carlo replications (Figure~1 of the main text). Informative censoring refers to censoring with a linear predictor. Relative bias and relative RMSE are expressed as a percentage of the true NB. Coverage refers to the 90\% bootstrap confidence intervals, and Monte Carlo standard errors are given in parentheses. \textit{Abbreviations:} CIPCW = conditional inverse probability of censoring weighting, Emp.\ SE = empirical standard error, KM = Kaplan--Meier, MCSE = Monte Carlo standard error, NB = net benefit, Rel. = relative, RMSE = root mean squared error, Thr. = decision threshold.}\label{tab:std_main}\\
\toprule
Thr. & True NB & Method & \makecell[r]{Rel.\ bias, \%\\(MCSE)} & \makecell[r]{Rel.\ RMSE, \%\\(MCSE)} & \makecell[r]{Bias\\(MCSE)} & \makecell[r]{Emp.\ SE\\(MCSE)} & \makecell[r]{RMSE\\(MCSE)} & \makecell[r]{Coverage, \%\\(MCSE)}\\
\midrule\endfirsthead
\multicolumn{9}{@{}l}{\textit{Table \thetable\ (continued)}}\\
\toprule
Thr. & True NB & Method & \makecell[r]{Rel.\ bias, \%\\(MCSE)} & \makecell[r]{Rel.\ RMSE, \%\\(MCSE)} & \makecell[r]{Bias\\(MCSE)} & \makecell[r]{Emp.\ SE\\(MCSE)} & \makecell[r]{RMSE\\(MCSE)} & \makecell[r]{Coverage, \%\\(MCSE)}\\
\midrule\endhead
\bottomrule\endlastfoot
\addlinespace[2pt]\multicolumn{9}{@{}l}{\textbf{Non-informative censoring, 10\% censored}}\\*
1\% & 0.2927 & KM & 0.4 (0.3) & 8.5 (0.2) & 0.0011 (0.0008) & 0.0250 (0.0006) & 0.0250 (0.0005) & 89.8 (1.0)\\
 &  & CIPCW (Cox) & 0.0 (0.3) & 8.5 (0.2) & 0.0001 (0.0008) & 0.0250 (0.0006) & 0.0250 (0.0005) & 89.3 (1.0)\\
 &  & CIPCW (Misspecified Cox) & 0.0 (0.3) & 8.5 (0.2) & 0.0001 (0.0008) & 0.0250 (0.0006) & 0.0250 (0.0005) & 89.6 (1.0)\\
10\% & 0.2445 & KM & 0.4 (0.3) & 10.5 (0.2) & 0.0010 (0.0008) & 0.0256 (0.0006) & 0.0257 (0.0006) & 89.8 (1.0)\\
 &  & CIPCW (Cox) & 0.0 (0.3) & 10.5 (0.2) & 0.0001 (0.0008) & 0.0256 (0.0006) & 0.0256 (0.0006) & 89.1 (1.0)\\
 &  & CIPCW (Misspecified Cox) & 0.0 (0.3) & 10.5 (0.2) & 0.0001 (0.0008) & 0.0256 (0.0006) & 0.0256 (0.0006) & 89.0 (1.0)\\
25\% & 0.1911 & KM & 0.5 (0.4) & 13.5 (0.3) & 0.0009 (0.0008) & 0.0257 (0.0006) & 0.0257 (0.0006) & 89.5 (1.0)\\
 &  & CIPCW (Cox) & 0.0 (0.4) & 13.4 (0.3) & 0.0001 (0.0008) & 0.0257 (0.0006) & 0.0257 (0.0006) & 90.2 (0.9)\\
 &  & CIPCW (Misspecified Cox) & 0.0 (0.4) & 13.4 (0.3) & 0.0001 (0.0008) & 0.0257 (0.0006) & 0.0257 (0.0006) & 90.0 (0.9)\\
50\% & 0.1308 & KM & 0.7 (0.6) & 19.6 (0.4) & 0.0009 (0.0008) & 0.0256 (0.0006) & 0.0256 (0.0005) & 90.0 (0.9)\\
 &  & CIPCW (Cox) & 0.1 (0.6) & 19.6 (0.4) & 0.0001 (0.0008) & 0.0257 (0.0006) & 0.0257 (0.0006) & 90.1 (0.9)\\
 &  & CIPCW (Misspecified Cox) & 0.1 (0.6) & 19.6 (0.4) & 0.0001 (0.0008) & 0.0257 (0.0006) & 0.0256 (0.0005) & 90.0 (0.9)\\
75\% & 0.0836 & KM & 0.1 (1.0) & 31.1 (0.7) & 0.0001 (0.0008) & 0.0260 (0.0006) & 0.0260 (0.0006) & 89.1 (1.0)\\
 &  & CIPCW (Cox) & $-$0.1 (1.0) & 31.1 (0.7) & $-$0.0001 (0.0008) & 0.0260 (0.0006) & 0.0260 (0.0006) & 88.9 (1.0)\\
 &  & CIPCW (Misspecified Cox) & $-$0.1 (1.0) & 31.1 (0.7) & $-$0.0001 (0.0008) & 0.0260 (0.0006) & 0.0260 (0.0006) & 89.0 (1.0)\\
\addlinespace[2pt]\multicolumn{9}{@{}l}{\textbf{Non-informative censoring, 20\% censored}}\\*
1\% & 0.2927 & KM & 0.4 (0.3) & 8.6 (0.2) & 0.0011 (0.0008) & 0.0252 (0.0006) & 0.0252 (0.0006) & 90.2 (0.9)\\
 &  & CIPCW (Cox) & 0.1 (0.3) & 8.5 (0.2) & 0.0002 (0.0008) & 0.0250 (0.0006) & 0.0250 (0.0006) & 90.1 (0.9)\\
 &  & CIPCW (Misspecified Cox) & 0.1 (0.3) & 8.6 (0.2) & 0.0002 (0.0008) & 0.0251 (0.0006) & 0.0251 (0.0006) & 90.2 (0.9)\\
10\% & 0.2445 & KM & 0.3 (0.3) & 10.5 (0.2) & 0.0008 (0.0008) & 0.0258 (0.0006) & 0.0258 (0.0006) & 89.8 (1.0)\\
 &  & CIPCW (Cox) & 0.0 (0.3) & 10.5 (0.2) & 0.0000 (0.0008) & 0.0257 (0.0006) & 0.0257 (0.0006) & 90.0 (0.9)\\
 &  & CIPCW (Misspecified Cox) & 0.0 (0.3) & 10.5 (0.2) & 0.0000 (0.0008) & 0.0258 (0.0006) & 0.0258 (0.0006) & 89.9 (1.0)\\
25\% & 0.1911 & KM & 0.5 (0.4) & 13.7 (0.3) & 0.0010 (0.0008) & 0.0262 (0.0006) & 0.0262 (0.0006) & 89.2 (1.0)\\
 &  & CIPCW (Cox) & 0.2 (0.4) & 13.6 (0.3) & 0.0004 (0.0008) & 0.0260 (0.0006) & 0.0260 (0.0006) & 89.6 (1.0)\\
 &  & CIPCW (Misspecified Cox) & 0.2 (0.4) & 13.6 (0.3) & 0.0004 (0.0008) & 0.0260 (0.0006) & 0.0260 (0.0006) & 89.8 (1.0)\\
50\% & 0.1306 & KM & 0.5 (0.6) & 20.0 (0.4) & 0.0007 (0.0008) & 0.0261 (0.0006) & 0.0261 (0.0006) & 90.5 (0.9)\\
 &  & CIPCW (Cox) & 0.3 (0.6) & 20.1 (0.4) & 0.0004 (0.0008) & 0.0262 (0.0006) & 0.0262 (0.0006) & 90.4 (0.9)\\
 &  & CIPCW (Misspecified Cox) & 0.3 (0.6) & 20.1 (0.4) & 0.0004 (0.0008) & 0.0262 (0.0006) & 0.0262 (0.0006) & 90.5 (0.9)\\
75\% & 0.0835 & KM & 0.5 (1.0) & 31.1 (0.7) & 0.0004 (0.0008) & 0.0260 (0.0006) & 0.0260 (0.0006) & 87.8 (1.0)\\
 &  & CIPCW (Cox) & 0.5 (1.0) & 31.6 (0.7) & 0.0004 (0.0008) & 0.0264 (0.0006) & 0.0264 (0.0006) & 87.6 (1.0)\\
 &  & CIPCW (Misspecified Cox) & 0.5 (1.0) & 31.6 (0.7) & 0.0005 (0.0008) & 0.0264 (0.0006) & 0.0264 (0.0006) & 87.6 (1.0)\\
\addlinespace[2pt]\multicolumn{9}{@{}l}{\textbf{Non-informative censoring, 30\% censored}}\\*
1\% & 0.2927 & KM & 0.5 (0.3) & 8.8 (0.2) & 0.0016 (0.0008) & 0.0259 (0.0006) & 0.0259 (0.0006) & 90.2 (0.9)\\
 &  & CIPCW (Cox) & 0.2 (0.3) & 8.7 (0.2) & 0.0007 (0.0008) & 0.0255 (0.0006) & 0.0255 (0.0006) & 89.5 (1.0)\\
 &  & CIPCW (Misspecified Cox) & 0.3 (0.3) & 8.8 (0.2) & 0.0007 (0.0008) & 0.0256 (0.0006) & 0.0256 (0.0006) & 89.9 (1.0)\\
10\% & 0.2447 & KM & 0.5 (0.3) & 10.8 (0.2) & 0.0012 (0.0008) & 0.0264 (0.0006) & 0.0264 (0.0006) & 90.4 (0.9)\\
 &  & CIPCW (Cox) & 0.1 (0.3) & 10.7 (0.2) & 0.0003 (0.0008) & 0.0263 (0.0006) & 0.0263 (0.0006) & 90.2 (0.9)\\
 &  & CIPCW (Misspecified Cox) & 0.1 (0.3) & 10.8 (0.2) & 0.0003 (0.0008) & 0.0264 (0.0006) & 0.0264 (0.0006) & 90.4 (0.9)\\
25\% & 0.1912 & KM & 0.6 (0.4) & 13.8 (0.3) & 0.0012 (0.0008) & 0.0263 (0.0006) & 0.0263 (0.0006) & 88.6 (1.0)\\
 &  & CIPCW (Cox) & 0.2 (0.4) & 13.7 (0.3) & 0.0003 (0.0008) & 0.0261 (0.0006) & 0.0261 (0.0006) & 89.8 (1.0)\\
 &  & CIPCW (Misspecified Cox) & 0.2 (0.4) & 13.7 (0.3) & 0.0003 (0.0008) & 0.0262 (0.0006) & 0.0262 (0.0006) & 89.8 (1.0)\\
50\% & 0.1309 & KM & 0.8 (0.7) & 20.6 (0.5) & 0.0011 (0.0009) & 0.0270 (0.0006) & 0.0270 (0.0006) & 88.4 (1.0)\\
 &  & CIPCW (Cox) & 0.2 (0.7) & 20.8 (0.5) & 0.0002 (0.0009) & 0.0273 (0.0006) & 0.0272 (0.0006) & 88.6 (1.0)\\
 &  & CIPCW (Misspecified Cox) & 0.2 (0.7) & 20.8 (0.5) & 0.0003 (0.0009) & 0.0272 (0.0006) & 0.0272 (0.0006) & 88.8 (1.0)\\
75\% & 0.0836 & KM & 0.3 (1.1) & 34.0 (0.8) & 0.0003 (0.0009) & 0.0285 (0.0006) & 0.0284 (0.0007) & 86.9 (1.1)\\
 &  & CIPCW (Cox) & $-$0.3 (1.1) & 34.6 (0.8) & $-$0.0002 (0.0009) & 0.0290 (0.0006) & 0.0290 (0.0007) & 86.8 (1.1)\\
 &  & CIPCW (Misspecified Cox) & $-$0.3 (1.1) & 34.6 (0.8) & $-$0.0002 (0.0009) & 0.0289 (0.0006) & 0.0289 (0.0007) & 87.0 (1.1)\\
\addlinespace[2pt]\multicolumn{9}{@{}l}{\textbf{Linear informative censoring, 10\% censored}}\\*
1\% & 0.2927 & KM & $-$9.1 (0.3) & 12.1 (0.2) & $-$0.0265 (0.0007) & 0.0237 (0.0005) & 0.0355 (0.0006) & 70.7 (1.4)\\
 &  & CIPCW (Cox) & 0.0 (0.3) & 8.4 (0.2) & 0.0000 (0.0008) & 0.0247 (0.0006) & 0.0247 (0.0005) & 89.0 (1.0)\\
 &  & CIPCW (Misspecified Cox) & $-$4.7 (0.3) & 9.6 (0.2) & $-$0.0139 (0.0008) & 0.0243 (0.0005) & 0.0280 (0.0006) & 83.9 (1.2)\\
10\% & 0.2445 & KM & $-$8.6 (0.3) & 13.3 (0.3) & $-$0.0209 (0.0008) & 0.0248 (0.0006) & 0.0325 (0.0006) & 78.2 (1.3)\\
 &  & CIPCW (Cox) & $-$0.1 (0.3) & 10.4 (0.2) & $-$0.0002 (0.0008) & 0.0255 (0.0006) & 0.0255 (0.0005) & 90.0 (0.9)\\
 &  & CIPCW (Misspecified Cox) & $-$5.9 (0.3) & 11.8 (0.2) & $-$0.0145 (0.0008) & 0.0249 (0.0006) & 0.0288 (0.0006) & 85.2 (1.1)\\
25\% & 0.1911 & KM & $-$8.1 (0.4) & 15.8 (0.3) & $-$0.0155 (0.0008) & 0.0259 (0.0006) & 0.0302 (0.0006) & 84.3 (1.2)\\
 &  & CIPCW (Cox) & $-$0.1 (0.4) & 13.7 (0.3) & $-$0.0002 (0.0008) & 0.0262 (0.0006) & 0.0262 (0.0006) & 88.5 (1.0)\\
 &  & CIPCW (Misspecified Cox) & $-$7.0 (0.4) & 14.9 (0.3) & $-$0.0133 (0.0008) & 0.0252 (0.0006) & 0.0285 (0.0006) & 84.9 (1.1)\\
50\% & 0.1308 & KM & $-$7.8 (0.7) & 22.5 (0.5) & $-$0.0102 (0.0009) & 0.0276 (0.0006) & 0.0294 (0.0006) & 86.8 (1.1)\\
 &  & CIPCW (Cox) & $-$0.2 (0.7) & 21.1 (0.5) & $-$0.0002 (0.0009) & 0.0276 (0.0006) & 0.0275 (0.0006) & 89.5 (1.0)\\
 &  & CIPCW (Misspecified Cox) & $-$7.1 (0.6) & 20.3 (0.4) & $-$0.0093 (0.0008) & 0.0248 (0.0006) & 0.0265 (0.0005) & 87.5 (1.0)\\
75\% & 0.0836 & KM & $-$8.5 (1.2) & 40.0 (1.0) & $-$0.0071 (0.0010) & 0.0327 (0.0007) & 0.0334 (0.0008) & 87.9 (1.0)\\
 &  & CIPCW (Cox) & $-$0.1 (1.2) & 38.0 (1.0) & $-$0.0001 (0.0010) & 0.0318 (0.0007) & 0.0318 (0.0008) & 86.9 (1.1)\\
 &  & CIPCW (Misspecified Cox) & $-$5.0 (0.9) & 29.7 (0.8) & $-$0.0042 (0.0008) & 0.0245 (0.0005) & 0.0249 (0.0006) & 87.9 (1.0)\\
\addlinespace[2pt]\multicolumn{9}{@{}l}{\textbf{Linear informative censoring, 20\% censored}}\\*
1\% & 0.2927 & KM & $-$18.9 (0.2) & 20.2 (0.2) & $-$0.0552 (0.0007) & 0.0214 (0.0005) & 0.0593 (0.0007) & 20.3 (1.3)\\
 &  & CIPCW (Cox) & $-$0.2 (0.3) & 9.2 (0.2) & $-$0.0005 (0.0008) & 0.0268 (0.0006) & 0.0268 (0.0007) & 90.2 (0.9)\\
 &  & CIPCW (Misspecified Cox) & $-$10.0 (0.3) & 12.9 (0.2) & $-$0.0293 (0.0008) & 0.0238 (0.0005) & 0.0377 (0.0007) & 67.2 (1.5)\\
10\% & 0.2445 & KM & $-$18.3 (0.3) & 20.6 (0.3) & $-$0.0447 (0.0007) & 0.0233 (0.0005) & 0.0504 (0.0007) & 41.3 (1.6)\\
 &  & CIPCW (Cox) & $-$0.3 (0.4) & 11.5 (0.3) & $-$0.0007 (0.0009) & 0.0280 (0.0006) & 0.0280 (0.0008) & 91.3 (0.9)\\
 &  & CIPCW (Misspecified Cox) & $-$12.5 (0.3) & 16.0 (0.3) & $-$0.0304 (0.0008) & 0.0246 (0.0006) & 0.0391 (0.0007) & 65.2 (1.5)\\
25\% & 0.1911 & KM & $-$17.7 (0.4) & 22.2 (0.4) & $-$0.0339 (0.0008) & 0.0255 (0.0006) & 0.0424 (0.0007) & 62.8 (1.5)\\
 &  & CIPCW (Cox) & $-$0.2 (0.5) & 15.5 (0.4) & $-$0.0004 (0.0009) & 0.0296 (0.0007) & 0.0296 (0.0008) & 90.2 (0.9)\\
 &  & CIPCW (Misspecified Cox) & $-$14.4 (0.4) & 19.4 (0.3) & $-$0.0274 (0.0008) & 0.0248 (0.0006) & 0.0370 (0.0007) & 69.8 (1.5)\\
50\% & 0.1306 & KM & $-$16.6 (0.8) & 29.3 (0.6) & $-$0.0216 (0.0010) & 0.0315 (0.0007) & 0.0382 (0.0008) & 80.7 (1.2)\\
 &  & CIPCW (Cox) & 0.7 (0.8) & 26.3 (0.7) & 0.0009 (0.0011) & 0.0344 (0.0008) & 0.0344 (0.0009) & 87.7 (1.0)\\
 &  & CIPCW (Misspecified Cox) & $-$14.1 (0.6) & 24.4 (0.5) & $-$0.0184 (0.0008) & 0.0260 (0.0006) & 0.0319 (0.0007) & 78.9 (1.3)\\
75\% & 0.0835 & KM & $-$19.0 (1.7) & 57.6 (1.4) & $-$0.0159 (0.0014) & 0.0454 (0.0010) & 0.0481 (0.0012) & 85.1 (1.1)\\
 &  & CIPCW (Cox) & 3.4 (1.6) & 50.7 (1.3) & 0.0028 (0.0013) & 0.0423 (0.0009) & 0.0424 (0.0011) & 79.8 (1.3)\\
 &  & CIPCW (Misspecified Cox) & $-$11.5 (1.0) & 33.1 (2.1) & $-$0.0096 (0.0008) & 0.0260 (0.0006) & 0.0277 (0.0017) & 86.6 (1.1)\\
\addlinespace[2pt]\multicolumn{9}{@{}l}{\textbf{Linear informative censoring, 30\% censored}}\\*
1\% & 0.2927 & KM & $-$28.9 (0.2) & 29.7 (0.2) & $-$0.0845 (0.0006) & 0.0198 (0.0004) & 0.0868 (0.0006) & 0.6 (0.2)\\
 &  & CIPCW (Cox) & $-$1.8 (0.4) & 12.1 (0.3) & $-$0.0053 (0.0011) & 0.0350 (0.0008) & 0.0354 (0.0009) & 87.0 (1.1)\\
 &  & CIPCW (Misspecified Cox) & $-$15.8 (0.3) & 18.4 (0.3) & $-$0.0462 (0.0009) & 0.0275 (0.0006) & 0.0538 (0.0007) & 41.7 (1.6)\\
10\% & 0.2447 & KM & $-$28.6 (0.3) & 30.0 (0.3) & $-$0.0699 (0.0007) & 0.0228 (0.0005) & 0.0735 (0.0007) & 8.6 (0.9)\\
 &  & CIPCW (Cox) & $-$2.3 (0.5) & 15.2 (0.4) & $-$0.0056 (0.0012) & 0.0368 (0.0008) & 0.0372 (0.0009) & 86.3 (1.1)\\
 &  & CIPCW (Misspecified Cox) & $-$19.5 (0.4) & 22.7 (0.3) & $-$0.0478 (0.0009) & 0.0285 (0.0006) & 0.0557 (0.0008) & 41.6 (1.6)\\
25\% & 0.1912 & KM & $-$28.6 (0.4) & 31.9 (0.4) & $-$0.0548 (0.0008) & 0.0267 (0.0006) & 0.0609 (0.0008) & 34.9 (1.5)\\
 &  & CIPCW (Cox) & $-$2.7 (0.6) & 20.4 (0.5) & $-$0.0052 (0.0012) & 0.0387 (0.0009) & 0.0391 (0.0009) & 85.7 (1.1)\\
 &  & CIPCW (Misspecified Cox) & $-$22.4 (0.5) & 27.1 (0.4) & $-$0.0429 (0.0009) & 0.0291 (0.0007) & 0.0518 (0.0007) & 47.4 (1.6)\\
50\% & 0.1309 & KM & $-$28.6 (1.0) & 41.6 (0.8) & $-$0.0374 (0.0013) & 0.0396 (0.0009) & 0.0544 (0.0011) & 70.3 (1.4)\\
 &  & CIPCW (Cox) & $-$2.4 (1.1) & 35.7 (0.9) & $-$0.0031 (0.0015) & 0.0467 (0.0010) & 0.0467 (0.0012) & 87.5 (1.0)\\
 &  & CIPCW (Misspecified Cox) & $-$22.1 (0.8) & 32.9 (0.7) & $-$0.0289 (0.0010) & 0.0320 (0.0007) & 0.0431 (0.0010) & 63.9 (1.5)\\
75\% & 0.0836 & KM & $-$45.0 (2.6) & 95.0 (2.1) & $-$0.0376 (0.0022) & 0.0700 (0.0016) & 0.0794 (0.0018) & 76.6 (1.3)\\
 &  & CIPCW (Cox) & 3.7 (2.5) & 78.7 (3.4) & 0.0031 (0.0021) & 0.0658 (0.0015) & 0.0658 (0.0028) & 77.8 (1.3)\\
 &  & CIPCW (Misspecified Cox) & $-$20.4 (1.2) & 44.4 (3.4) & $-$0.0171 (0.0010) & 0.0330 (0.0007) & 0.0371 (0.0028) & 72.9 (1.4)\\
\end{longtable}

\begin{longtable}{@{}r r l r r r r r r@{}}
\caption{Performance of the net benefit estimators in the competing-risk setting at 10\%, 20\% and 30\% censoring, based on 1,000 Monte Carlo replications (Figure~2 of the main text). Informative censoring refers to censoring with a linear predictor. Relative bias and relative RMSE are expressed as a percentage of the true NB. Coverage refers to the 90\% bootstrap confidence intervals, and Monte Carlo standard errors are given in parentheses. \textit{Abbreviations:} AJ = Aalen--Johansen, CIPCW = conditional inverse probability of censoring weighting, Emp.\ SE = empirical standard error, MCSE = Monte Carlo standard error, NB = net benefit, Rel. = relative, RMSE = root mean squared error, Thr. = decision threshold.}\label{tab:cr_main}\\
\toprule
Thr. & True NB & Method & \makecell[r]{Rel.\ bias, \%\\(MCSE)} & \makecell[r]{Rel.\ RMSE, \%\\(MCSE)} & \makecell[r]{Bias\\(MCSE)} & \makecell[r]{Emp.\ SE\\(MCSE)} & \makecell[r]{RMSE\\(MCSE)} & \makecell[r]{Coverage, \%\\(MCSE)}\\
\midrule\endfirsthead
\multicolumn{9}{@{}l}{\textit{Table \thetable\ (continued)}}\\
\toprule
Thr. & True NB & Method & \makecell[r]{Rel.\ bias, \%\\(MCSE)} & \makecell[r]{Rel.\ RMSE, \%\\(MCSE)} & \makecell[r]{Bias\\(MCSE)} & \makecell[r]{Emp.\ SE\\(MCSE)} & \makecell[r]{RMSE\\(MCSE)} & \makecell[r]{Coverage, \%\\(MCSE)}\\
\midrule\endhead
\bottomrule\endlastfoot
\addlinespace[2pt]\multicolumn{9}{@{}l}{\textbf{Non-informative censoring, 10\% censored}}\\*
1\% & 0.2928 & AJ & 0.2 (0.3) & 8.7 (0.2) & 0.0007 (0.0008) & 0.0253 (0.0006) & 0.0253 (0.0006) & 89.5 (1.0)\\
 &  & CIPCW (Cox) & 0.2 (0.3) & 8.6 (0.2) & 0.0007 (0.0008) & 0.0252 (0.0006) & 0.0252 (0.0006) & 89.7 (1.0)\\
 &  & CIPCW (Misspecified Cox) & 0.2 (0.3) & 8.6 (0.2) & 0.0007 (0.0008) & 0.0252 (0.0006) & 0.0252 (0.0006) & 89.9 (1.0)\\
10\% & 0.2438 & AJ & 0.3 (0.3) & 10.7 (0.2) & 0.0008 (0.0008) & 0.0260 (0.0006) & 0.0260 (0.0006) & 89.4 (1.0)\\
 &  & CIPCW (Cox) & 0.3 (0.3) & 10.7 (0.2) & 0.0008 (0.0008) & 0.0260 (0.0006) & 0.0260 (0.0006) & 89.5 (1.0)\\
 &  & CIPCW (Misspecified Cox) & 0.3 (0.3) & 10.7 (0.2) & 0.0008 (0.0008) & 0.0260 (0.0006) & 0.0260 (0.0006) & 89.4 (1.0)\\
25\% & 0.1886 & AJ & 0.4 (0.4) & 13.5 (0.3) & 0.0008 (0.0008) & 0.0255 (0.0006) & 0.0255 (0.0006) & 90.2 (0.9)\\
 &  & CIPCW (Cox) & 0.5 (0.4) & 13.6 (0.3) & 0.0009 (0.0008) & 0.0256 (0.0006) & 0.0256 (0.0006) & 90.2 (0.9)\\
 &  & CIPCW (Misspecified Cox) & 0.4 (0.4) & 13.6 (0.3) & 0.0008 (0.0008) & 0.0256 (0.0006) & 0.0256 (0.0006) & 90.1 (0.9)\\
50\% & 0.1246 & AJ & 0.8 (0.6) & 20.5 (0.4) & 0.0010 (0.0008) & 0.0256 (0.0006) & 0.0256 (0.0006) & 88.6 (1.0)\\
 &  & CIPCW (Cox) & 0.8 (0.7) & 20.6 (0.5) & 0.0010 (0.0008) & 0.0257 (0.0006) & 0.0257 (0.0006) & 89.2 (1.0)\\
 &  & CIPCW (Misspecified Cox) & 0.8 (0.7) & 20.6 (0.5) & 0.0010 (0.0008) & 0.0257 (0.0006) & 0.0257 (0.0006) & 89.1 (1.0)\\
75\% & 0.0726 & AJ & 0.7 (1.1) & 35.2 (0.8) & 0.0005 (0.0008) & 0.0255 (0.0006) & 0.0255 (0.0006) & 88.4 (1.0)\\
 &  & CIPCW (Cox) & 0.5 (1.1) & 35.3 (0.8) & 0.0004 (0.0008) & 0.0257 (0.0006) & 0.0257 (0.0006) & 88.6 (1.0)\\
 &  & CIPCW (Misspecified Cox) & 0.5 (1.1) & 35.4 (0.8) & 0.0004 (0.0008) & 0.0257 (0.0006) & 0.0257 (0.0006) & 88.4 (1.0)\\
\addlinespace[2pt]\multicolumn{9}{@{}l}{\textbf{Non-informative censoring, 20\% censored}}\\*
1\% & 0.2929 & AJ & 0.1 (0.3) & 8.7 (0.2) & 0.0004 (0.0008) & 0.0254 (0.0006) & 0.0254 (0.0006) & 89.5 (1.0)\\
 &  & CIPCW (Cox) & 0.1 (0.3) & 8.6 (0.2) & 0.0003 (0.0008) & 0.0252 (0.0006) & 0.0252 (0.0005) & 89.7 (1.0)\\
 &  & CIPCW (Misspecified Cox) & 0.1 (0.3) & 8.6 (0.2) & 0.0002 (0.0008) & 0.0253 (0.0006) & 0.0252 (0.0005) & 89.8 (1.0)\\
10\% & 0.2439 & AJ & 0.2 (0.3) & 10.6 (0.2) & 0.0005 (0.0008) & 0.0259 (0.0006) & 0.0259 (0.0006) & 89.9 (1.0)\\
 &  & CIPCW (Cox) & 0.2 (0.3) & 10.6 (0.2) & 0.0004 (0.0008) & 0.0259 (0.0006) & 0.0259 (0.0006) & 89.6 (1.0)\\
 &  & CIPCW (Misspecified Cox) & 0.1 (0.3) & 10.6 (0.2) & 0.0004 (0.0008) & 0.0260 (0.0006) & 0.0259 (0.0006) & 89.7 (1.0)\\
25\% & 0.1886 & AJ & 0.3 (0.4) & 13.5 (0.3) & 0.0006 (0.0008) & 0.0254 (0.0006) & 0.0254 (0.0006) & 90.3 (0.9)\\
 &  & CIPCW (Cox) & 0.3 (0.4) & 13.6 (0.3) & 0.0006 (0.0008) & 0.0256 (0.0006) & 0.0256 (0.0006) & 90.1 (0.9)\\
 &  & CIPCW (Misspecified Cox) & 0.3 (0.4) & 13.6 (0.3) & 0.0006 (0.0008) & 0.0256 (0.0006) & 0.0256 (0.0006) & 90.0 (0.9)\\
50\% & 0.1247 & AJ & 0.6 (0.7) & 20.6 (0.5) & 0.0008 (0.0008) & 0.0256 (0.0006) & 0.0256 (0.0006) & 88.9 (1.0)\\
 &  & CIPCW (Cox) & 0.5 (0.7) & 20.7 (0.5) & 0.0007 (0.0008) & 0.0258 (0.0006) & 0.0258 (0.0006) & 88.8 (1.0)\\
 &  & CIPCW (Misspecified Cox) & 0.5 (0.7) & 20.6 (0.5) & 0.0007 (0.0008) & 0.0258 (0.0006) & 0.0257 (0.0006) & 88.6 (1.0)\\
75\% & 0.0728 & AJ & 0.5 (1.1) & 36.0 (0.8) & 0.0004 (0.0008) & 0.0262 (0.0006) & 0.0262 (0.0006) & 87.2 (1.1)\\
 &  & CIPCW (Cox) & 0.3 (1.1) & 35.8 (0.8) & 0.0002 (0.0008) & 0.0261 (0.0006) & 0.0261 (0.0006) & 87.8 (1.0)\\
 &  & CIPCW (Misspecified Cox) & 0.3 (1.1) & 35.8 (0.8) & 0.0002 (0.0008) & 0.0261 (0.0006) & 0.0261 (0.0006) & 87.9 (1.0)\\
\addlinespace[2pt]\multicolumn{9}{@{}l}{\textbf{Non-informative censoring, 30\% censored}}\\*
1\% & 0.2928 & AJ & 0.3 (0.3) & 9.0 (0.2) & 0.0008 (0.0008) & 0.0264 (0.0006) & 0.0264 (0.0006) & 88.6 (1.0)\\
 &  & CIPCW (Cox) & 0.2 (0.3) & 8.8 (0.2) & 0.0006 (0.0008) & 0.0259 (0.0006) & 0.0259 (0.0006) & 88.3 (1.0)\\
 &  & CIPCW (Misspecified Cox) & 0.2 (0.3) & 8.9 (0.2) & 0.0006 (0.0008) & 0.0261 (0.0006) & 0.0260 (0.0006) & 88.3 (1.0)\\
10\% & 0.2439 & AJ & 0.3 (0.3) & 11.1 (0.3) & 0.0007 (0.0009) & 0.0270 (0.0006) & 0.0270 (0.0006) & 88.6 (1.0)\\
 &  & CIPCW (Cox) & 0.2 (0.3) & 11.0 (0.2) & 0.0004 (0.0008) & 0.0268 (0.0006) & 0.0268 (0.0006) & 89.1 (1.0)\\
 &  & CIPCW (Misspecified Cox) & 0.2 (0.3) & 11.0 (0.2) & 0.0004 (0.0008) & 0.0269 (0.0006) & 0.0269 (0.0006) & 88.8 (1.0)\\
25\% & 0.1886 & AJ & 0.3 (0.4) & 13.7 (0.3) & 0.0005 (0.0008) & 0.0259 (0.0006) & 0.0259 (0.0006) & 89.4 (1.0)\\
 &  & CIPCW (Cox) & 0.2 (0.4) & 13.7 (0.3) & 0.0003 (0.0008) & 0.0259 (0.0006) & 0.0259 (0.0006) & 89.9 (1.0)\\
 &  & CIPCW (Misspecified Cox) & 0.2 (0.4) & 13.8 (0.3) & 0.0003 (0.0008) & 0.0259 (0.0006) & 0.0259 (0.0006) & 90.0 (0.9)\\
50\% & 0.1248 & AJ & 0.5 (0.7) & 20.8 (0.5) & 0.0006 (0.0008) & 0.0259 (0.0006) & 0.0259 (0.0006) & 89.5 (1.0)\\
 &  & CIPCW (Cox) & 0.4 (0.7) & 20.9 (0.4) & 0.0005 (0.0008) & 0.0260 (0.0006) & 0.0260 (0.0006) & 89.3 (1.0)\\
 &  & CIPCW (Misspecified Cox) & 0.4 (0.7) & 20.8 (0.4) & 0.0005 (0.0008) & 0.0260 (0.0006) & 0.0260 (0.0006) & 89.6 (1.0)\\
75\% & 0.0727 & AJ & 0.6 (1.1) & 35.9 (0.8) & 0.0004 (0.0008) & 0.0261 (0.0006) & 0.0261 (0.0006) & 89.4 (1.0)\\
 &  & CIPCW (Cox) & 0.1 (1.2) & 36.7 (0.9) & 0.0000 (0.0008) & 0.0267 (0.0006) & 0.0267 (0.0006) & 89.1 (1.0)\\
 &  & CIPCW (Misspecified Cox) & 0.1 (1.2) & 36.7 (0.9) & 0.0001 (0.0008) & 0.0267 (0.0006) & 0.0267 (0.0006) & 88.5 (1.0)\\
\addlinespace[2pt]\multicolumn{9}{@{}l}{\textbf{Linear informative censoring, 10\% censored}}\\*
1\% & 0.2928 & AJ & $-$9.5 (0.3) & 12.4 (0.2) & $-$0.0277 (0.0008) & 0.0237 (0.0005) & 0.0364 (0.0006) & 66.7 (1.5)\\
 &  & CIPCW (Cox) & 0.0 (0.3) & 8.4 (0.2) & 0.0001 (0.0008) & 0.0246 (0.0005) & 0.0246 (0.0005) & 90.1 (0.9)\\
 &  & CIPCW (Misspecified Cox) & $-$3.9 (0.3) & 9.2 (0.2) & $-$0.0114 (0.0008) & 0.0244 (0.0005) & 0.0269 (0.0006) & 85.6 (1.1)\\
10\% & 0.2438 & AJ & $-$8.9 (0.3) & 13.6 (0.3) & $-$0.0218 (0.0008) & 0.0250 (0.0006) & 0.0331 (0.0006) & 76.9 (1.3)\\
 &  & CIPCW (Cox) & 0.0 (0.3) & 10.5 (0.2) & 0.0000 (0.0008) & 0.0255 (0.0006) & 0.0255 (0.0006) & 89.9 (1.0)\\
 &  & CIPCW (Misspecified Cox) & $-$4.9 (0.3) & 11.4 (0.2) & $-$0.0119 (0.0008) & 0.0253 (0.0006) & 0.0279 (0.0006) & 85.6 (1.1)\\
25\% & 0.1886 & AJ & $-$8.6 (0.4) & 15.8 (0.3) & $-$0.0163 (0.0008) & 0.0250 (0.0006) & 0.0298 (0.0006) & 83.1 (1.2)\\
 &  & CIPCW (Cox) & $-$0.1 (0.4) & 13.5 (0.3) & $-$0.0002 (0.0008) & 0.0254 (0.0006) & 0.0254 (0.0006) & 90.1 (0.9)\\
 &  & CIPCW (Misspecified Cox) & $-$6.0 (0.4) & 14.4 (0.3) & $-$0.0113 (0.0008) & 0.0247 (0.0006) & 0.0271 (0.0006) & 86.5 (1.1)\\
50\% & 0.1246 & AJ & $-$8.0 (0.7) & 23.4 (0.5) & $-$0.0100 (0.0009) & 0.0274 (0.0006) & 0.0292 (0.0006) & 87.4 (1.0)\\
 &  & CIPCW (Cox) & 0.0 (0.7) & 21.6 (0.5) & 0.0000 (0.0009) & 0.0269 (0.0006) & 0.0269 (0.0006) & 89.6 (1.0)\\
 &  & CIPCW (Misspecified Cox) & $-$6.5 (0.6) & 21.1 (0.5) & $-$0.0081 (0.0008) & 0.0251 (0.0006) & 0.0264 (0.0006) & 88.0 (1.0)\\
75\% & 0.0726 & AJ & $-$8.9 (1.4) & 46.0 (1.1) & $-$0.0065 (0.0010) & 0.0328 (0.0007) & 0.0334 (0.0008) & 88.1 (1.0)\\
 &  & CIPCW (Cox) & $-$0.7 (1.4) & 42.8 (1.1) & $-$0.0005 (0.0010) & 0.0311 (0.0007) & 0.0310 (0.0008) & 87.2 (1.1)\\
 &  & CIPCW (Misspecified Cox) & $-$6.7 (1.1) & 36.4 (0.9) & $-$0.0049 (0.0008) & 0.0260 (0.0006) & 0.0264 (0.0007) & 88.0 (1.0)\\
\addlinespace[2pt]\multicolumn{9}{@{}l}{\textbf{Linear informative censoring, 20\% censored}}\\*
1\% & 0.2929 & AJ & $-$19.1 (0.2) & 20.5 (0.2) & $-$0.0560 (0.0007) & 0.0218 (0.0005) & 0.0601 (0.0007) & 18.6 (1.2)\\
 &  & CIPCW (Cox) & $-$0.1 (0.3) & 9.3 (0.2) & $-$0.0002 (0.0009) & 0.0274 (0.0006) & 0.0273 (0.0006) & 90.0 (0.9)\\
 &  & CIPCW (Misspecified Cox) & $-$8.3 (0.3) & 12.0 (0.2) & $-$0.0243 (0.0008) & 0.0256 (0.0006) & 0.0353 (0.0007) & 73.3 (1.4)\\
10\% & 0.2439 & AJ & $-$18.4 (0.3) & 20.8 (0.3) & $-$0.0448 (0.0008) & 0.0239 (0.0005) & 0.0507 (0.0007) & 41.0 (1.6)\\
 &  & CIPCW (Cox) & $-$0.1 (0.4) & 11.8 (0.3) & $-$0.0004 (0.0009) & 0.0288 (0.0006) & 0.0288 (0.0007) & 89.3 (1.0)\\
 &  & CIPCW (Misspecified Cox) & $-$10.4 (0.3) & 15.2 (0.3) & $-$0.0255 (0.0008) & 0.0268 (0.0006) & 0.0370 (0.0007) & 72.7 (1.4)\\
25\% & 0.1886 & AJ & $-$17.9 (0.4) & 22.4 (0.4) & $-$0.0337 (0.0008) & 0.0256 (0.0006) & 0.0423 (0.0007) & 62.7 (1.5)\\
 &  & CIPCW (Cox) & $-$0.3 (0.5) & 15.8 (0.4) & $-$0.0005 (0.0009) & 0.0299 (0.0007) & 0.0298 (0.0007) & 90.6 (0.9)\\
 &  & CIPCW (Misspecified Cox) & $-$12.8 (0.5) & 19.2 (0.4) & $-$0.0241 (0.0009) & 0.0270 (0.0006) & 0.0362 (0.0007) & 72.9 (1.4)\\
50\% & 0.1247 & AJ & $-$15.8 (0.8) & 29.8 (0.6) & $-$0.0197 (0.0010) & 0.0315 (0.0007) & 0.0371 (0.0007) & 81.0 (1.2)\\
 &  & CIPCW (Cox) & 0.9 (0.8) & 26.6 (0.6) & 0.0011 (0.0010) & 0.0332 (0.0007) & 0.0332 (0.0008) & 88.9 (1.0)\\
 &  & CIPCW (Misspecified Cox) & $-$13.7 (0.7) & 26.7 (0.6) & $-$0.0170 (0.0009) & 0.0286 (0.0006) & 0.0333 (0.0008) & 80.9 (1.2)\\
75\% & 0.0728 & AJ & $-$20.6 (1.9) & 64.8 (1.5) & $-$0.0150 (0.0014) & 0.0447 (0.0010) & 0.0471 (0.0011) & 84.0 (1.2)\\
 &  & CIPCW (Cox) & 3.4 (1.9) & 58.6 (2.6) & 0.0025 (0.0013) & 0.0426 (0.0010) & 0.0426 (0.0019) & 80.8 (1.2)\\
 &  & CIPCW (Misspecified Cox) & $-$13.0 (1.3) & 43.5 (2.2) & $-$0.0094 (0.0010) & 0.0302 (0.0007) & 0.0316 (0.0016) & 88.1 (1.0)\\
\addlinespace[2pt]\multicolumn{9}{@{}l}{\textbf{Linear informative censoring, 30\% censored}}\\*
1\% & 0.2928 & AJ & $-$29.2 (0.2) & 29.9 (0.2) & $-$0.0854 (0.0006) & 0.0199 (0.0004) & 0.0877 (0.0006) & 0.7 (0.3)\\
 &  & CIPCW (Cox) & $-$2.5 (0.4) & 11.5 (0.3) & $-$0.0072 (0.0010) & 0.0328 (0.0007) & 0.0336 (0.0007) & 86.5 (1.1)\\
 &  & CIPCW (Misspecified Cox) & $-$13.7 (0.3) & 17.0 (0.3) & $-$0.0402 (0.0009) & 0.0292 (0.0007) & 0.0497 (0.0007) & 52.1 (1.6)\\
10\% & 0.2439 & AJ & $-$28.6 (0.3) & 30.1 (0.3) & $-$0.0697 (0.0007) & 0.0228 (0.0005) & 0.0734 (0.0007) & 9.1 (0.9)\\
 &  & CIPCW (Cox) & $-$3.1 (0.4) & 14.5 (0.3) & $-$0.0076 (0.0011) & 0.0346 (0.0008) & 0.0354 (0.0008) & 86.7 (1.1)\\
 &  & CIPCW (Misspecified Cox) & $-$17.3 (0.4) & 21.3 (0.3) & $-$0.0421 (0.0010) & 0.0307 (0.0007) & 0.0521 (0.0008) & 53.2 (1.6)\\
25\% & 0.1886 & AJ & $-$28.5 (0.5) & 31.9 (0.4) & $-$0.0537 (0.0009) & 0.0270 (0.0006) & 0.0602 (0.0008) & 36.9 (1.5)\\
 &  & CIPCW (Cox) & $-$3.9 (0.6) & 19.9 (0.4) & $-$0.0074 (0.0012) & 0.0368 (0.0008) & 0.0375 (0.0008) & 87.0 (1.1)\\
 &  & CIPCW (Misspecified Cox) & $-$20.9 (0.5) & 26.7 (0.4) & $-$0.0394 (0.0010) & 0.0315 (0.0007) & 0.0504 (0.0008) & 58.1 (1.6)\\
50\% & 0.1248 & AJ & $-$28.2 (1.0) & 42.9 (0.8) & $-$0.0352 (0.0013) & 0.0403 (0.0009) & 0.0535 (0.0010) & 75.3 (1.4)\\
 &  & CIPCW (Cox) & $-$3.9 (1.1) & 35.3 (0.9) & $-$0.0048 (0.0014) & 0.0438 (0.0010) & 0.0440 (0.0011) & 88.4 (1.0)\\
 &  & CIPCW (Misspecified Cox) & $-$22.7 (0.9) & 35.9 (0.8) & $-$0.0284 (0.0011) & 0.0346 (0.0008) & 0.0448 (0.0010) & 67.7 (1.5)\\
75\% & 0.0727 & AJ & $-$57.5 (2.8) & 106.6 (2.4) & $-$0.0418 (0.0021) & 0.0653 (0.0015) & 0.0775 (0.0017) & 76.9 (1.3)\\
 &  & CIPCW (Cox) & $-$0.5 (2.7) & 85.2 (3.8) & $-$0.0004 (0.0020) & 0.0620 (0.0014) & 0.0619 (0.0027) & 84.9 (1.1)\\
 &  & CIPCW (Misspecified Cox) & $-$21.7 (1.4) & 50.6 (2.8) & $-$0.0157 (0.0011) & 0.0333 (0.0007) & 0.0368 (0.0021) & 75.6 (1.4)\\
\end{longtable}
}

\clearpage
\section{Sensitivity analysis}\label{sec:sens}

\begin{figure}[H]
\centering
\includegraphics[width=\textwidth,height=0.72\textheight,keepaspectratio]{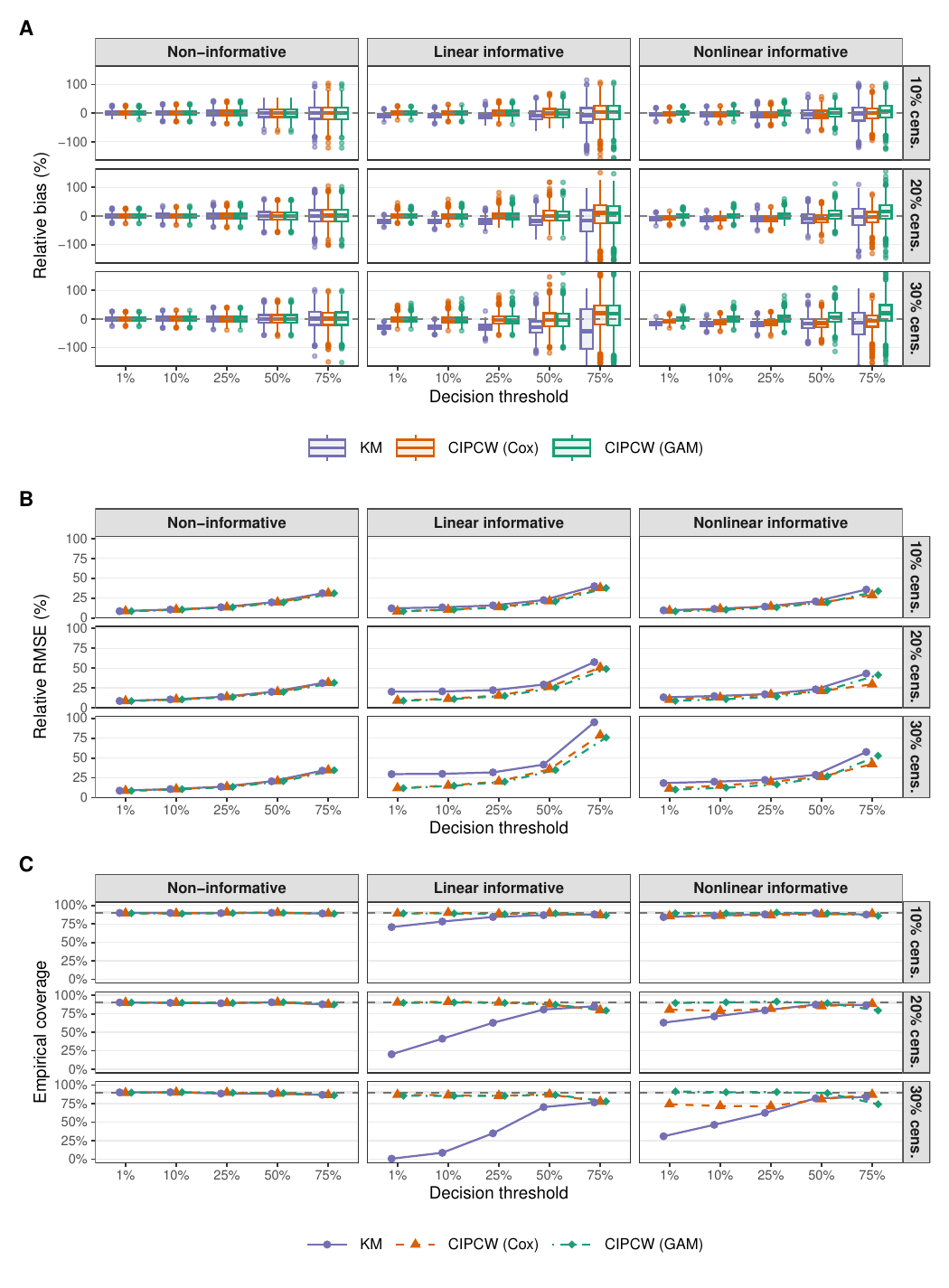}
\caption{Sensitivity analysis for the standard survival setting at 10\%, 20\% and 30\% censoring, based on 1,000 Monte Carlo replications. (A) Distribution of the relative bias in the estimated NB (estimated minus true NB, as a percentage of the true NB) across decision thresholds. The vertical axis is limited to $\pm$150\%. (B) Relative root mean squared error (RMSE) of the estimated NB, as a percentage of the true NB. (C) Empirical coverage of the 90\% bootstrap confidence intervals. The dashed line marks the nominal level and the shaded band its Monte Carlo interval. CIPCW (Cox) and CIPCW (GAM) estimate the censoring weights using a Cox model with linear effects and a Cox generalized additive model (Cox GAM), respectively. \textit{Abbreviations:} CIPCW = conditional inverse probability of censoring weighting, GAM = generalized additive model, KM = Kaplan--Meier, NB = net benefit.}
\label{fig:std_sens}
\end{figure}

\begin{figure}[H]
\centering
\includegraphics[width=\textwidth,height=0.72\textheight,keepaspectratio]{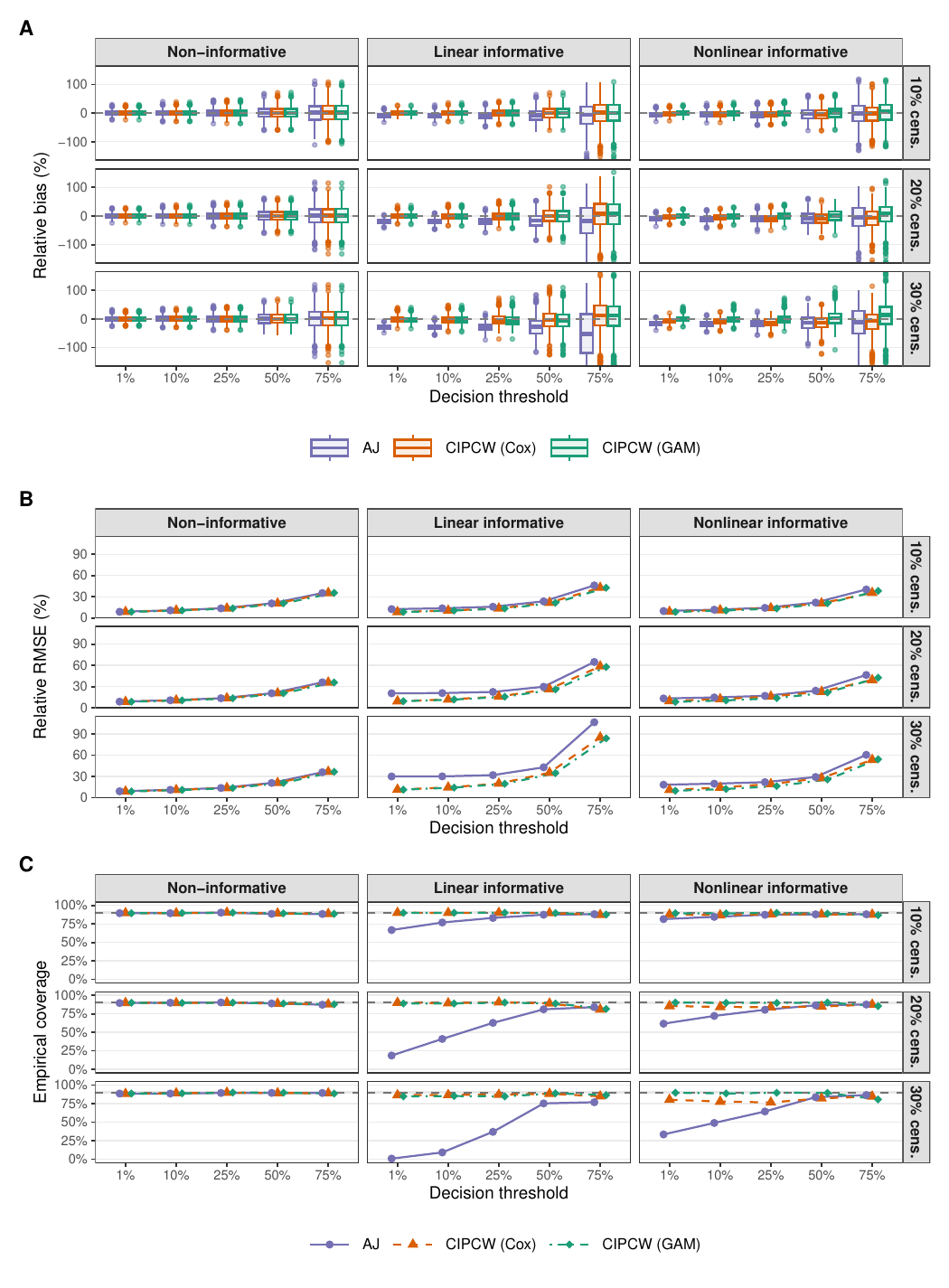}
\caption{Sensitivity analysis for the competing-risk setting at 10\%, 20\% and 30\% censoring, based on 1,000 Monte Carlo replications. (A) Distribution of the relative bias in the estimated NB (estimated minus true NB, as a percentage of the true NB) across decision thresholds. The vertical axis is limited to $\pm$150\%. (B) Relative root mean squared error (RMSE) of the estimated NB, as a percentage of the true NB. (C) Empirical coverage of the 90\% bootstrap confidence intervals. The dashed line marks the nominal level and the shaded band its Monte Carlo interval. CIPCW (Cox) and CIPCW (GAM) estimate the censoring weights using a Cox model with linear effects and a Cox generalized additive model (Cox GAM), respectively. \textit{Abbreviations:} AJ = Aalen--Johansen, CIPCW = conditional inverse probability of censoring weighting, GAM = generalized additive model, NB = net benefit.}
\label{fig:cr_sens}
\end{figure}

\FloatBarrier
{\scriptsize\setlength{\tabcolsep}{2.2pt}
\begin{longtable}{@{}r r l r r r r r r@{}}
\caption{Sensitivity analysis in the standard survival setting at 10\%, 20\% and 30\% censoring, based on 1,000 Monte Carlo replications (Figure~\ref{fig:std_sens}). Under nonlinear informative censoring, all approaches are shown. Under non-informative and linear informative censoring, only the CIPCW-based approach with the Cox GAM censoring model is shown, as the other approaches are reported in Table~\ref{tab:std_main}. Relative bias and relative RMSE are expressed as a percentage of the true NB. Coverage refers to the 90\% bootstrap confidence intervals, and Monte Carlo standard errors are given in parentheses. \textit{Abbreviations:} CIPCW = conditional inverse probability of censoring weighting, Emp.\ SE = empirical standard error, GAM = generalized additive model, KM = Kaplan--Meier, MCSE = Monte Carlo standard error, NB = net benefit, Rel. = relative, RMSE = root mean squared error, Thr. = decision threshold.}\label{tab:std_sens}\\
\toprule
Thr. & True NB & Method & \makecell[r]{Rel.\ bias, \%\\(MCSE)} & \makecell[r]{Rel.\ RMSE, \%\\(MCSE)} & \makecell[r]{Bias\\(MCSE)} & \makecell[r]{Emp.\ SE\\(MCSE)} & \makecell[r]{RMSE\\(MCSE)} & \makecell[r]{Coverage, \%\\(MCSE)}\\
\midrule\endfirsthead
\multicolumn{9}{@{}l}{\textit{Table \thetable\ (continued)}}\\
\toprule
Thr. & True NB & Method & \makecell[r]{Rel.\ bias, \%\\(MCSE)} & \makecell[r]{Rel.\ RMSE, \%\\(MCSE)} & \makecell[r]{Bias\\(MCSE)} & \makecell[r]{Emp.\ SE\\(MCSE)} & \makecell[r]{RMSE\\(MCSE)} & \makecell[r]{Coverage, \%\\(MCSE)}\\
\midrule\endhead
\bottomrule\endlastfoot
\addlinespace[2pt]\multicolumn{9}{@{}l}{\textbf{Nonlinear informative censoring, 10\% censored}}\\*
1\% & 0.2927 & KM & $-$4.8 (0.3) & 9.6 (0.2) & $-$0.0142 (0.0008) & 0.0242 (0.0005) & 0.0281 (0.0006) & 84.2 (1.2)\\
 &  & CIPCW (Cox) & $-$3.4 (0.3) & 9.1 (0.2) & $-$0.0100 (0.0008) & 0.0246 (0.0006) & 0.0266 (0.0006) & 85.6 (1.1)\\
 &  & CIPCW (GAM) & $-$0.3 (0.3) & 8.3 (0.2) & $-$0.0008 (0.0008) & 0.0244 (0.0005) & 0.0244 (0.0005) & 89.3 (1.0)\\
10\% & 0.2445 & KM & $-$4.9 (0.3) & 11.4 (0.2) & $-$0.0119 (0.0008) & 0.0252 (0.0006) & 0.0278 (0.0006) & 86.2 (1.1)\\
 &  & CIPCW (Cox) & $-$4.4 (0.3) & 11.2 (0.2) & $-$0.0108 (0.0008) & 0.0252 (0.0006) & 0.0274 (0.0006) & 86.2 (1.1)\\
 &  & CIPCW (GAM) & $-$0.4 (0.3) & 10.3 (0.2) & $-$0.0010 (0.0008) & 0.0251 (0.0006) & 0.0251 (0.0006) & 89.2 (1.0)\\
25\% & 0.1911 & KM & $-$4.7 (0.4) & 14.3 (0.3) & $-$0.0089 (0.0008) & 0.0258 (0.0006) & 0.0273 (0.0006) & 87.9 (1.0)\\
 &  & CIPCW (Cox) & $-$5.4 (0.4) & 14.3 (0.3) & $-$0.0102 (0.0008) & 0.0253 (0.0006) & 0.0273 (0.0006) & 86.7 (1.1)\\
 &  & CIPCW (GAM) & $-$0.2 (0.4) & 13.3 (0.3) & $-$0.0004 (0.0008) & 0.0255 (0.0006) & 0.0255 (0.0006) & 90.2 (0.9)\\
50\% & 0.1308 & KM & $-$3.8 (0.6) & 20.9 (0.5) & $-$0.0050 (0.0008) & 0.0268 (0.0006) & 0.0273 (0.0006) & 89.6 (1.0)\\
 &  & CIPCW (Cox) & $-$5.0 (0.6) & 19.5 (0.4) & $-$0.0066 (0.0008) & 0.0247 (0.0006) & 0.0256 (0.0005) & 88.0 (1.0)\\
 &  & CIPCW (GAM) & 1.3 (0.6) & 19.8 (0.4) & 0.0017 (0.0008) & 0.0259 (0.0006) & 0.0259 (0.0006) & 89.1 (1.0)\\
75\% & 0.0836 & KM & $-$3.0 (1.1) & 35.7 (0.9) & $-$0.0025 (0.0009) & 0.0298 (0.0007) & 0.0298 (0.0007) & 87.6 (1.0)\\
 &  & CIPCW (Cox) & $-$2.1 (0.9) & 28.8 (0.7) & $-$0.0017 (0.0008) & 0.0240 (0.0005) & 0.0241 (0.0006) & 88.5 (1.0)\\
 &  & CIPCW (GAM) & 4.4 (1.1) & 33.8 (1.0) & 0.0037 (0.0009) & 0.0280 (0.0006) & 0.0282 (0.0008) & 85.9 (1.1)\\
\addlinespace[2pt]\multicolumn{9}{@{}l}{\textbf{Nonlinear informative censoring, 20\% censored}}\\*
1\% & 0.2927 & KM & $-$10.6 (0.2) & 13.2 (0.2) & $-$0.0310 (0.0007) & 0.0231 (0.0005) & 0.0387 (0.0007) & 63.0 (1.5)\\
 &  & CIPCW (Cox) & $-$5.9 (0.3) & 10.2 (0.2) & $-$0.0174 (0.0008) & 0.0244 (0.0005) & 0.0299 (0.0006) & 80.6 (1.3)\\
 &  & CIPCW (GAM) & 0.1 (0.3) & 8.7 (0.2) & 0.0003 (0.0008) & 0.0255 (0.0006) & 0.0255 (0.0006) & 89.5 (1.0)\\
10\% & 0.2445 & KM & $-$11.0 (0.3) & 14.8 (0.3) & $-$0.0269 (0.0008) & 0.0242 (0.0005) & 0.0362 (0.0007) & 71.4 (1.4)\\
 &  & CIPCW (Cox) & $-$7.9 (0.3) & 13.0 (0.3) & $-$0.0194 (0.0008) & 0.0251 (0.0006) & 0.0317 (0.0007) & 79.0 (1.3)\\
 &  & CIPCW (GAM) & 0.0 (0.3) & 10.8 (0.3) & $-$0.0001 (0.0008) & 0.0264 (0.0006) & 0.0264 (0.0007) & 90.5 (0.9)\\
25\% & 0.1911 & KM & $-$10.7 (0.4) & 17.1 (0.3) & $-$0.0204 (0.0008) & 0.0254 (0.0006) & 0.0326 (0.0006) & 79.6 (1.3)\\
 &  & CIPCW (Cox) & $-$9.9 (0.4) & 16.5 (0.3) & $-$0.0189 (0.0008) & 0.0253 (0.0006) & 0.0316 (0.0006) & 81.6 (1.2)\\
 &  & CIPCW (GAM) & 0.5 (0.4) & 14.1 (0.3) & 0.0009 (0.0009) & 0.0270 (0.0006) & 0.0270 (0.0006) & 91.5 (0.9)\\
50\% & 0.1306 & KM & $-$8.6 (0.7) & 23.3 (0.5) & $-$0.0113 (0.0009) & 0.0282 (0.0006) & 0.0304 (0.0006) & 87.4 (1.0)\\
 &  & CIPCW (Cox) & $-$9.4 (0.6) & 21.3 (0.5) & $-$0.0122 (0.0008) & 0.0250 (0.0006) & 0.0278 (0.0006) & 85.4 (1.1)\\
 &  & CIPCW (GAM) & 4.4 (0.7) & 22.3 (0.6) & 0.0057 (0.0009) & 0.0286 (0.0006) & 0.0292 (0.0008) & 89.4 (1.0)\\
75\% & 0.0835 & KM & $-$5.8 (1.4) & 43.2 (1.0) & $-$0.0049 (0.0011) & 0.0358 (0.0008) & 0.0361 (0.0008) & 86.7 (1.1)\\
 &  & CIPCW (Cox) & $-$4.8 (0.9) & 29.4 (1.3) & $-$0.0040 (0.0008) & 0.0243 (0.0005) & 0.0246 (0.0011) & 88.3 (1.0)\\
 &  & CIPCW (GAM) & 12.5 (1.2) & 41.4 (1.3) & 0.0105 (0.0010) & 0.0329 (0.0007) & 0.0345 (0.0011) & 79.5 (1.3)\\
\addlinespace[2pt]\multicolumn{9}{@{}l}{\textbf{Nonlinear informative censoring, 30\% censored}}\\*
1\% & 0.2927 & KM & $-$16.8 (0.2) & 18.4 (0.2) & $-$0.0493 (0.0007) & 0.0218 (0.0005) & 0.0539 (0.0007) & 30.9 (1.5)\\
 &  & CIPCW (Cox) & $-$8.0 (0.3) & 11.6 (0.2) & $-$0.0234 (0.0008) & 0.0248 (0.0006) & 0.0341 (0.0006) & 74.2 (1.4)\\
 &  & CIPCW (GAM) & 0.1 (0.3) & 10.1 (0.3) & 0.0002 (0.0009) & 0.0295 (0.0007) & 0.0295 (0.0008) & 91.5 (0.9)\\
10\% & 0.2447 & KM & $-$17.7 (0.3) & 20.1 (0.3) & $-$0.0432 (0.0007) & 0.0236 (0.0005) & 0.0492 (0.0007) & 46.4 (1.6)\\
 &  & CIPCW (Cox) & $-$10.9 (0.3) & 15.2 (0.3) & $-$0.0267 (0.0008) & 0.0260 (0.0006) & 0.0372 (0.0007) & 71.7 (1.4)\\
 &  & CIPCW (GAM) & $-$0.1 (0.4) & 12.6 (0.3) & $-$0.0002 (0.0010) & 0.0308 (0.0007) & 0.0308 (0.0008) & 90.4 (0.9)\\
25\% & 0.1912 & KM & $-$17.9 (0.4) & 22.3 (0.4) & $-$0.0341 (0.0008) & 0.0256 (0.0006) & 0.0427 (0.0007) & 62.5 (1.5)\\
 &  & CIPCW (Cox) & $-$14.1 (0.4) & 19.9 (0.4) & $-$0.0269 (0.0009) & 0.0269 (0.0006) & 0.0381 (0.0007) & 71.4 (1.4)\\
 &  & CIPCW (GAM) & 0.5 (0.5) & 16.6 (0.5) & 0.0010 (0.0010) & 0.0318 (0.0007) & 0.0318 (0.0009) & 90.7 (0.9)\\
50\% & 0.1309 & KM & $-$14.9 (0.8) & 28.8 (0.6) & $-$0.0194 (0.0010) & 0.0323 (0.0007) & 0.0377 (0.0008) & 82.2 (1.2)\\
 &  & CIPCW (Cox) & $-$14.1 (0.7) & 26.2 (0.6) & $-$0.0184 (0.0009) & 0.0290 (0.0006) & 0.0343 (0.0008) & 81.1 (1.2)\\
 &  & CIPCW (GAM) & 6.5 (0.8) & 27.0 (0.7) & 0.0086 (0.0011) & 0.0343 (0.0008) & 0.0353 (0.0010) & 89.1 (1.0)\\
75\% & 0.0836 & KM & $-$16.0 (1.8) & 57.6 (1.4) & $-$0.0134 (0.0015) & 0.0463 (0.0010) & 0.0482 (0.0011) & 84.0 (1.2)\\
 &  & CIPCW (Cox) & $-$12.1 (1.3) & 42.3 (2.6) & $-$0.0101 (0.0011) & 0.0339 (0.0008) & 0.0354 (0.0022) & 87.2 (1.1)\\
 &  & CIPCW (GAM) & 17.7 (1.6) & 52.8 (2.0) & 0.0148 (0.0013) & 0.0416 (0.0009) & 0.0441 (0.0016) & 74.2 (1.4)\\
\addlinespace[2pt]\multicolumn{9}{@{}l}{\textbf{Non-informative censoring, 10\% censored}}\\*
1\% & 0.2927 & CIPCW (GAM) & 0.0 (0.3) & 8.5 (0.2) & 0.0001 (0.0008) & 0.0249 (0.0006) & 0.0249 (0.0005) & 89.4 (1.0)\\
10\% & 0.2445 & CIPCW (GAM) & 0.0 (0.3) & 10.4 (0.2) & 0.0001 (0.0008) & 0.0256 (0.0006) & 0.0255 (0.0006) & 88.9 (1.0)\\
25\% & 0.1911 & CIPCW (GAM) & 0.0 (0.4) & 13.4 (0.3) & 0.0001 (0.0008) & 0.0257 (0.0006) & 0.0257 (0.0006) & 89.6 (1.0)\\
50\% & 0.1308 & CIPCW (GAM) & 0.1 (0.6) & 19.6 (0.4) & 0.0002 (0.0008) & 0.0257 (0.0006) & 0.0257 (0.0006) & 89.9 (1.0)\\
75\% & 0.0836 & CIPCW (GAM) & $-$0.1 (1.0) & 31.2 (0.7) & $-$0.0001 (0.0008) & 0.0261 (0.0006) & 0.0260 (0.0006) & 88.8 (1.0)\\
\addlinespace[2pt]\multicolumn{9}{@{}l}{\textbf{Non-informative censoring, 20\% censored}}\\*
1\% & 0.2927 & CIPCW (GAM) & 0.0 (0.3) & 8.5 (0.2) & 0.0001 (0.0008) & 0.0249 (0.0006) & 0.0249 (0.0005) & 90.2 (0.9)\\
10\% & 0.2445 & CIPCW (GAM) & 0.0 (0.3) & 10.5 (0.2) & $-$0.0001 (0.0008) & 0.0256 (0.0006) & 0.0256 (0.0006) & 89.9 (1.0)\\
25\% & 0.1911 & CIPCW (GAM) & 0.2 (0.4) & 13.6 (0.3) & 0.0003 (0.0008) & 0.0259 (0.0006) & 0.0259 (0.0006) & 89.6 (1.0)\\
50\% & 0.1306 & CIPCW (GAM) & 0.3 (0.6) & 20.1 (0.4) & 0.0004 (0.0008) & 0.0262 (0.0006) & 0.0262 (0.0006) & 90.7 (0.9)\\
75\% & 0.0835 & CIPCW (GAM) & 0.6 (1.0) & 31.6 (0.7) & 0.0005 (0.0008) & 0.0264 (0.0006) & 0.0264 (0.0006) & 87.4 (1.0)\\
\addlinespace[2pt]\multicolumn{9}{@{}l}{\textbf{Non-informative censoring, 30\% censored}}\\*
1\% & 0.2927 & CIPCW (GAM) & 0.2 (0.3) & 8.7 (0.2) & 0.0006 (0.0008) & 0.0254 (0.0006) & 0.0254 (0.0005) & 89.9 (1.0)\\
10\% & 0.2447 & CIPCW (GAM) & 0.1 (0.3) & 10.7 (0.2) & 0.0002 (0.0008) & 0.0261 (0.0006) & 0.0261 (0.0006) & 90.4 (0.9)\\
25\% & 0.1912 & CIPCW (GAM) & 0.1 (0.4) & 13.6 (0.3) & 0.0002 (0.0008) & 0.0260 (0.0006) & 0.0260 (0.0006) & 89.7 (1.0)\\
50\% & 0.1309 & CIPCW (GAM) & 0.2 (0.7) & 20.8 (0.5) & 0.0003 (0.0009) & 0.0272 (0.0006) & 0.0272 (0.0006) & 89.2 (1.0)\\
75\% & 0.0836 & CIPCW (GAM) & $-$0.2 (1.1) & 34.6 (0.8) & $-$0.0002 (0.0009) & 0.0289 (0.0006) & 0.0289 (0.0007) & 86.3 (1.1)\\
\addlinespace[2pt]\multicolumn{9}{@{}l}{\textbf{Linear informative censoring, 10\% censored}}\\*
1\% & 0.2927 & CIPCW (GAM) & $-$0.2 (0.3) & 8.4 (0.2) & $-$0.0004 (0.0008) & 0.0246 (0.0005) & 0.0246 (0.0005) & 89.1 (1.0)\\
10\% & 0.2445 & CIPCW (GAM) & $-$0.3 (0.3) & 10.4 (0.2) & $-$0.0006 (0.0008) & 0.0253 (0.0006) & 0.0253 (0.0005) & 89.1 (1.0)\\
25\% & 0.1911 & CIPCW (GAM) & $-$0.3 (0.4) & 13.6 (0.3) & $-$0.0006 (0.0008) & 0.0260 (0.0006) & 0.0260 (0.0006) & 88.3 (1.0)\\
50\% & 0.1308 & CIPCW (GAM) & $-$0.5 (0.7) & 20.8 (0.4) & $-$0.0006 (0.0009) & 0.0272 (0.0006) & 0.0272 (0.0006) & 89.3 (1.0)\\
75\% & 0.0836 & CIPCW (GAM) & $-$0.5 (1.2) & 37.8 (1.0) & $-$0.0004 (0.0010) & 0.0316 (0.0007) & 0.0316 (0.0008) & 86.5 (1.1)\\
\addlinespace[2pt]\multicolumn{9}{@{}l}{\textbf{Linear informative censoring, 20\% censored}}\\*
1\% & 0.2927 & CIPCW (GAM) & $-$0.5 (0.3) & 8.9 (0.2) & $-$0.0015 (0.0008) & 0.0259 (0.0006) & 0.0259 (0.0006) & 89.7 (1.0)\\
10\% & 0.2445 & CIPCW (GAM) & $-$0.8 (0.3) & 11.1 (0.3) & $-$0.0018 (0.0009) & 0.0271 (0.0006) & 0.0271 (0.0007) & 90.3 (0.9)\\
25\% & 0.1911 & CIPCW (GAM) & $-$0.7 (0.5) & 15.0 (0.4) & $-$0.0014 (0.0009) & 0.0286 (0.0006) & 0.0286 (0.0007) & 89.6 (1.0)\\
50\% & 0.1306 & CIPCW (GAM) & 0.1 (0.8) & 25.4 (0.6) & 0.0001 (0.0010) & 0.0332 (0.0007) & 0.0332 (0.0008) & 87.3 (1.1)\\
75\% & 0.0835 & CIPCW (GAM) & 2.9 (1.5) & 48.9 (1.3) & 0.0025 (0.0013) & 0.0408 (0.0009) & 0.0409 (0.0011) & 79.3 (1.3)\\
\addlinespace[2pt]\multicolumn{9}{@{}l}{\textbf{Linear informative censoring, 30\% censored}}\\*
1\% & 0.2927 & CIPCW (GAM) & $-$2.1 (0.4) & 12.0 (0.3) & $-$0.0062 (0.0011) & 0.0344 (0.0008) & 0.0350 (0.0009) & 85.6 (1.1)\\
10\% & 0.2447 & CIPCW (GAM) & $-$2.7 (0.5) & 15.0 (0.4) & $-$0.0066 (0.0011) & 0.0362 (0.0008) & 0.0367 (0.0009) & 85.4 (1.1)\\
25\% & 0.1912 & CIPCW (GAM) & $-$3.1 (0.6) & 20.1 (0.5) & $-$0.0060 (0.0012) & 0.0379 (0.0008) & 0.0384 (0.0010) & 85.6 (1.1)\\
50\% & 0.1309 & CIPCW (GAM) & $-$2.8 (1.1) & 34.8 (0.9) & $-$0.0036 (0.0014) & 0.0455 (0.0010) & 0.0456 (0.0012) & 86.8 (1.1)\\
75\% & 0.0836 & CIPCW (GAM) & 3.6 (2.4) & 75.7 (3.0) & 0.0030 (0.0020) & 0.0633 (0.0014) & 0.0633 (0.0026) & 78.0 (1.3)\\
\end{longtable}

\begin{longtable}{@{}r r l r r r r r r@{}}
\caption{Sensitivity analysis in the competing-risk setting at 10\%, 20\% and 30\% censoring, based on 1,000 Monte Carlo replications (Figure~\ref{fig:cr_sens}). Under nonlinear informative censoring, all approaches are shown. Under non-informative and linear informative censoring, only the CIPCW-based approach with the Cox GAM censoring model is shown, as the other approaches are reported in Table~\ref{tab:cr_main}. Relative bias and relative RMSE are expressed as a percentage of the true NB. Coverage refers to the 90\% bootstrap confidence intervals, and Monte Carlo standard errors are given in parentheses. \textit{Abbreviations:} AJ = Aalen--Johansen, CIPCW = conditional inverse probability of censoring weighting, Emp.\ SE = empirical standard error, GAM = generalized additive model, MCSE = Monte Carlo standard error, NB = net benefit, Rel. = relative, RMSE = root mean squared error, Thr. = decision threshold.}\label{tab:cr_sens}\\
\toprule
Thr. & True NB & Method & \makecell[r]{Rel.\ bias, \%\\(MCSE)} & \makecell[r]{Rel.\ RMSE, \%\\(MCSE)} & \makecell[r]{Bias\\(MCSE)} & \makecell[r]{Emp.\ SE\\(MCSE)} & \makecell[r]{RMSE\\(MCSE)} & \makecell[r]{Coverage, \%\\(MCSE)}\\
\midrule\endfirsthead
\multicolumn{9}{@{}l}{\textit{Table \thetable\ (continued)}}\\
\toprule
Thr. & True NB & Method & \makecell[r]{Rel.\ bias, \%\\(MCSE)} & \makecell[r]{Rel.\ RMSE, \%\\(MCSE)} & \makecell[r]{Bias\\(MCSE)} & \makecell[r]{Emp.\ SE\\(MCSE)} & \makecell[r]{RMSE\\(MCSE)} & \makecell[r]{Coverage, \%\\(MCSE)}\\
\midrule\endhead
\bottomrule\endlastfoot
\addlinespace[2pt]\multicolumn{9}{@{}l}{\textbf{Nonlinear informative censoring, 10\% censored}}\\*
1\% & 0.2928 & AJ & $-$4.9 (0.3) & 9.8 (0.2) & $-$0.0143 (0.0008) & 0.0248 (0.0006) & 0.0286 (0.0006) & 81.5 (1.2)\\
 &  & CIPCW (Cox) & $-$2.4 (0.3) & 8.9 (0.2) & $-$0.0069 (0.0008) & 0.0250 (0.0006) & 0.0260 (0.0006) & 87.6 (1.0)\\
 &  & CIPCW (GAM) & $-$0.3 (0.3) & 8.5 (0.2) & $-$0.0009 (0.0008) & 0.0249 (0.0006) & 0.0249 (0.0006) & 89.3 (1.0)\\
10\% & 0.2438 & AJ & $-$4.8 (0.3) & 11.6 (0.2) & $-$0.0117 (0.0008) & 0.0258 (0.0006) & 0.0283 (0.0006) & 84.5 (1.1)\\
 &  & CIPCW (Cox) & $-$3.1 (0.3) & 11.1 (0.2) & $-$0.0076 (0.0008) & 0.0259 (0.0006) & 0.0269 (0.0006) & 86.8 (1.1)\\
 &  & CIPCW (GAM) & $-$0.4 (0.3) & 10.6 (0.2) & $-$0.0010 (0.0008) & 0.0257 (0.0006) & 0.0257 (0.0006) & 89.3 (1.0)\\
25\% & 0.1886 & AJ & $-$4.6 (0.4) & 14.2 (0.3) & $-$0.0086 (0.0008) & 0.0254 (0.0006) & 0.0269 (0.0006) & 87.3 (1.1)\\
 &  & CIPCW (Cox) & $-$4.1 (0.4) & 14.0 (0.3) & $-$0.0077 (0.0008) & 0.0252 (0.0006) & 0.0264 (0.0006) & 88.0 (1.0)\\
 &  & CIPCW (GAM) & $-$0.5 (0.4) & 13.4 (0.3) & $-$0.0009 (0.0008) & 0.0252 (0.0006) & 0.0252 (0.0006) & 89.8 (1.0)\\
50\% & 0.1246 & AJ & $-$3.3 (0.7) & 21.7 (0.5) & $-$0.0041 (0.0008) & 0.0267 (0.0006) & 0.0270 (0.0006) & 87.9 (1.0)\\
 &  & CIPCW (Cox) & $-$4.0 (0.6) & 20.7 (0.4) & $-$0.0050 (0.0008) & 0.0253 (0.0006) & 0.0258 (0.0006) & 87.9 (1.0)\\
 &  & CIPCW (GAM) & 0.7 (0.7) & 20.8 (0.5) & 0.0009 (0.0008) & 0.0259 (0.0006) & 0.0259 (0.0006) & 88.9 (1.0)\\
75\% & 0.0726 & AJ & $-$2.5 (1.3) & 40.2 (0.9) & $-$0.0018 (0.0009) & 0.0292 (0.0007) & 0.0292 (0.0007) & 88.0 (1.0)\\
 &  & CIPCW (Cox) & $-$3.4 (1.1) & 35.4 (0.8) & $-$0.0025 (0.0008) & 0.0256 (0.0006) & 0.0257 (0.0006) & 87.7 (1.0)\\
 &  & CIPCW (GAM) & 2.2 (1.2) & 38.2 (0.9) & 0.0016 (0.0009) & 0.0277 (0.0006) & 0.0277 (0.0006) & 86.9 (1.1)\\
\addlinespace[2pt]\multicolumn{9}{@{}l}{\textbf{Nonlinear informative censoring, 20\% censored}}\\*
1\% & 0.2929 & AJ & $-$10.4 (0.3) & 13.2 (0.2) & $-$0.0305 (0.0008) & 0.0237 (0.0005) & 0.0386 (0.0007) & 61.6 (1.5)\\
 &  & CIPCW (Cox) & $-$4.4 (0.3) & 9.6 (0.2) & $-$0.0129 (0.0008) & 0.0251 (0.0006) & 0.0282 (0.0006) & 85.5 (1.1)\\
 &  & CIPCW (GAM) & $-$0.5 (0.3) & 8.4 (0.2) & $-$0.0016 (0.0008) & 0.0245 (0.0005) & 0.0246 (0.0005) & 90.1 (0.9)\\
10\% & 0.2439 & AJ & $-$10.4 (0.3) & 14.6 (0.3) & $-$0.0254 (0.0008) & 0.0250 (0.0006) & 0.0356 (0.0007) & 72.1 (1.4)\\
 &  & CIPCW (Cox) & $-$6.1 (0.3) & 12.3 (0.3) & $-$0.0148 (0.0008) & 0.0260 (0.0006) & 0.0299 (0.0006) & 84.3 (1.2)\\
 &  & CIPCW (GAM) & $-$0.8 (0.3) & 10.5 (0.2) & $-$0.0020 (0.0008) & 0.0255 (0.0006) & 0.0255 (0.0005) & 89.8 (1.0)\\
25\% & 0.1886 & AJ & $-$10.2 (0.4) & 16.9 (0.3) & $-$0.0193 (0.0008) & 0.0253 (0.0006) & 0.0319 (0.0006) & 80.4 (1.3)\\
 &  & CIPCW (Cox) & $-$8.4 (0.4) & 16.0 (0.3) & $-$0.0159 (0.0008) & 0.0257 (0.0006) & 0.0302 (0.0006) & 83.8 (1.2)\\
 &  & CIPCW (GAM) & $-$1.2 (0.4) & 13.7 (0.3) & $-$0.0022 (0.0008) & 0.0257 (0.0006) & 0.0258 (0.0006) & 89.6 (1.0)\\
50\% & 0.1247 & AJ & $-$7.7 (0.7) & 24.1 (0.5) & $-$0.0096 (0.0009) & 0.0285 (0.0006) & 0.0301 (0.0006) & 86.4 (1.1)\\
 &  & CIPCW (Cox) & $-$9.4 (0.7) & 22.8 (0.5) & $-$0.0117 (0.0008) & 0.0259 (0.0006) & 0.0284 (0.0006) & 84.8 (1.1)\\
 &  & CIPCW (GAM) & 0.9 (0.7) & 21.6 (0.5) & 0.0011 (0.0008) & 0.0269 (0.0006) & 0.0269 (0.0006) & 90.1 (0.9)\\
75\% & 0.0728 & AJ & $-$6.7 (1.5) & 46.4 (1.2) & $-$0.0049 (0.0011) & 0.0334 (0.0007) & 0.0338 (0.0008) & 87.5 (1.0)\\
 &  & CIPCW (Cox) & $-$9.1 (1.2) & 39.4 (1.6) & $-$0.0067 (0.0009) & 0.0279 (0.0006) & 0.0286 (0.0011) & 87.4 (1.0)\\
 &  & CIPCW (GAM) & 4.7 (1.3) & 42.5 (1.2) & 0.0034 (0.0010) & 0.0307 (0.0007) & 0.0309 (0.0009) & 85.4 (1.1)\\
\addlinespace[2pt]\multicolumn{9}{@{}l}{\textbf{Nonlinear informative censoring, 30\% censored}}\\*
1\% & 0.2928 & AJ & $-$16.6 (0.2) & 18.3 (0.2) & $-$0.0485 (0.0007) & 0.0227 (0.0005) & 0.0536 (0.0007) & 33.4 (1.5)\\
 &  & CIPCW (Cox) & $-$6.2 (0.3) & 10.9 (0.2) & $-$0.0181 (0.0008) & 0.0263 (0.0006) & 0.0319 (0.0007) & 80.3 (1.3)\\
 &  & CIPCW (GAM) & $-$0.8 (0.3) & 9.8 (0.3) & $-$0.0022 (0.0009) & 0.0285 (0.0006) & 0.0286 (0.0008) & 89.7 (1.0)\\
10\% & 0.2439 & AJ & $-$17.0 (0.3) & 19.8 (0.3) & $-$0.0415 (0.0008) & 0.0246 (0.0006) & 0.0482 (0.0007) & 48.9 (1.6)\\
 &  & CIPCW (Cox) & $-$8.7 (0.4) & 14.3 (0.3) & $-$0.0212 (0.0009) & 0.0277 (0.0006) & 0.0349 (0.0007) & 77.8 (1.3)\\
 &  & CIPCW (GAM) & $-$1.1 (0.4) & 12.3 (0.3) & $-$0.0028 (0.0009) & 0.0299 (0.0007) & 0.0300 (0.0008) & 88.7 (1.0)\\
25\% & 0.1886 & AJ & $-$16.9 (0.4) & 21.9 (0.4) & $-$0.0320 (0.0008) & 0.0262 (0.0006) & 0.0413 (0.0007) & 64.4 (1.5)\\
 &  & CIPCW (Cox) & $-$12.1 (0.5) & 19.3 (0.4) & $-$0.0229 (0.0009) & 0.0283 (0.0006) & 0.0364 (0.0007) & 76.6 (1.3)\\
 &  & CIPCW (GAM) & $-$1.6 (0.5) & 16.2 (0.4) & $-$0.0030 (0.0010) & 0.0305 (0.0007) & 0.0306 (0.0008) & 89.7 (1.0)\\
50\% & 0.1248 & AJ & $-$13.4 (0.8) & 29.2 (0.6) & $-$0.0167 (0.0010) & 0.0324 (0.0007) & 0.0364 (0.0008) & 83.7 (1.2)\\
 &  & CIPCW (Cox) & $-$13.8 (0.8) & 27.6 (0.7) & $-$0.0173 (0.0009) & 0.0299 (0.0007) & 0.0345 (0.0008) & 82.1 (1.2)\\
 &  & CIPCW (GAM) & 2.1 (0.8) & 25.7 (0.7) & 0.0027 (0.0010) & 0.0319 (0.0007) & 0.0320 (0.0009) & 89.4 (1.0)\\
75\% & 0.0727 & AJ & $-$15.1 (1.9) & 60.5 (1.6) & $-$0.0110 (0.0013) & 0.0426 (0.0010) & 0.0440 (0.0011) & 86.6 (1.1)\\
 &  & CIPCW (Cox) & $-$14.7 (1.6) & 53.6 (3.1) & $-$0.0107 (0.0012) & 0.0375 (0.0008) & 0.0390 (0.0023) & 84.8 (1.1)\\
 &  & CIPCW (GAM) & 10.7 (1.7) & 54.2 (1.8) & 0.0077 (0.0012) & 0.0387 (0.0009) & 0.0394 (0.0013) & 80.7 (1.2)\\
\addlinespace[2pt]\multicolumn{9}{@{}l}{\textbf{Non-informative censoring, 10\% censored}}\\*
1\% & 0.2928 & CIPCW (GAM) & 0.2 (0.3) & 8.6 (0.2) & 0.0006 (0.0008) & 0.0252 (0.0006) & 0.0252 (0.0006) & 89.5 (1.0)\\
10\% & 0.2438 & CIPCW (GAM) & 0.3 (0.3) & 10.6 (0.2) & 0.0007 (0.0008) & 0.0259 (0.0006) & 0.0259 (0.0006) & 89.7 (1.0)\\
25\% & 0.1886 & CIPCW (GAM) & 0.4 (0.4) & 13.6 (0.3) & 0.0008 (0.0008) & 0.0256 (0.0006) & 0.0256 (0.0006) & 90.1 (0.9)\\
50\% & 0.1246 & CIPCW (GAM) & 0.8 (0.7) & 20.6 (0.5) & 0.0010 (0.0008) & 0.0257 (0.0006) & 0.0257 (0.0006) & 89.2 (1.0)\\
75\% & 0.0726 & CIPCW (GAM) & 0.6 (1.1) & 35.4 (0.8) & 0.0004 (0.0008) & 0.0257 (0.0006) & 0.0257 (0.0006) & 88.6 (1.0)\\
\addlinespace[2pt]\multicolumn{9}{@{}l}{\textbf{Non-informative censoring, 20\% censored}}\\*
1\% & 0.2929 & CIPCW (GAM) & 0.1 (0.3) & 8.6 (0.2) & 0.0002 (0.0008) & 0.0251 (0.0006) & 0.0251 (0.0005) & 89.9 (1.0)\\
10\% & 0.2439 & CIPCW (GAM) & 0.1 (0.3) & 10.6 (0.2) & 0.0003 (0.0008) & 0.0258 (0.0006) & 0.0258 (0.0006) & 89.7 (1.0)\\
25\% & 0.1886 & CIPCW (GAM) & 0.3 (0.4) & 13.5 (0.3) & 0.0006 (0.0008) & 0.0255 (0.0006) & 0.0255 (0.0006) & 90.0 (0.9)\\
50\% & 0.1247 & CIPCW (GAM) & 0.5 (0.7) & 20.6 (0.5) & 0.0007 (0.0008) & 0.0257 (0.0006) & 0.0257 (0.0006) & 88.9 (1.0)\\
75\% & 0.0728 & CIPCW (GAM) & 0.3 (1.1) & 35.9 (0.8) & 0.0002 (0.0008) & 0.0261 (0.0006) & 0.0261 (0.0006) & 87.8 (1.0)\\
\addlinespace[2pt]\multicolumn{9}{@{}l}{\textbf{Non-informative censoring, 30\% censored}}\\*
1\% & 0.2928 & CIPCW (GAM) & 0.2 (0.3) & 8.8 (0.2) & 0.0005 (0.0008) & 0.0257 (0.0006) & 0.0257 (0.0006) & 88.7 (1.0)\\
10\% & 0.2439 & CIPCW (GAM) & 0.1 (0.3) & 10.9 (0.2) & 0.0003 (0.0008) & 0.0266 (0.0006) & 0.0265 (0.0006) & 88.9 (1.0)\\
25\% & 0.1886 & CIPCW (GAM) & 0.1 (0.4) & 13.7 (0.3) & 0.0002 (0.0008) & 0.0258 (0.0006) & 0.0257 (0.0006) & 89.9 (1.0)\\
50\% & 0.1248 & CIPCW (GAM) & 0.4 (0.7) & 20.8 (0.4) & 0.0005 (0.0008) & 0.0259 (0.0006) & 0.0259 (0.0006) & 89.8 (1.0)\\
75\% & 0.0727 & CIPCW (GAM) & 0.1 (1.2) & 36.6 (0.9) & 0.0001 (0.0008) & 0.0266 (0.0006) & 0.0266 (0.0006) & 88.5 (1.0)\\
\addlinespace[2pt]\multicolumn{9}{@{}l}{\textbf{Linear informative censoring, 10\% censored}}\\*
1\% & 0.2928 & CIPCW (GAM) & $-$0.2 (0.3) & 8.4 (0.2) & $-$0.0004 (0.0008) & 0.0245 (0.0005) & 0.0245 (0.0005) & 89.9 (1.0)\\
10\% & 0.2438 & CIPCW (GAM) & $-$0.2 (0.3) & 10.4 (0.2) & $-$0.0005 (0.0008) & 0.0254 (0.0006) & 0.0254 (0.0006) & 90.0 (0.9)\\
25\% & 0.1886 & CIPCW (GAM) & $-$0.4 (0.4) & 13.4 (0.3) & $-$0.0007 (0.0008) & 0.0252 (0.0006) & 0.0252 (0.0006) & 90.1 (0.9)\\
50\% & 0.1246 & CIPCW (GAM) & $-$0.4 (0.7) & 21.4 (0.5) & $-$0.0005 (0.0008) & 0.0267 (0.0006) & 0.0267 (0.0006) & 89.6 (1.0)\\
75\% & 0.0726 & CIPCW (GAM) & $-$1.3 (1.3) & 42.6 (1.1) & $-$0.0009 (0.0010) & 0.0310 (0.0007) & 0.0310 (0.0008) & 87.6 (1.0)\\
\addlinespace[2pt]\multicolumn{9}{@{}l}{\textbf{Linear informative censoring, 20\% censored}}\\*
1\% & 0.2929 & CIPCW (GAM) & $-$0.3 (0.3) & 9.3 (0.2) & $-$0.0010 (0.0009) & 0.0271 (0.0006) & 0.0271 (0.0006) & 88.8 (1.0)\\
10\% & 0.2439 & CIPCW (GAM) & $-$0.5 (0.4) & 11.7 (0.3) & $-$0.0011 (0.0009) & 0.0285 (0.0006) & 0.0285 (0.0006) & 88.9 (1.0)\\
25\% & 0.1886 & CIPCW (GAM) & $-$0.6 (0.5) & 15.6 (0.4) & $-$0.0012 (0.0009) & 0.0294 (0.0007) & 0.0294 (0.0007) & 90.1 (0.9)\\
50\% & 0.1247 & CIPCW (GAM) & 0.4 (0.8) & 26.1 (0.6) & 0.0005 (0.0010) & 0.0325 (0.0007) & 0.0325 (0.0008) & 88.8 (1.0)\\
75\% & 0.0728 & CIPCW (GAM) & 2.8 (1.8) & 57.7 (2.6) & 0.0020 (0.0013) & 0.0420 (0.0009) & 0.0420 (0.0019) & 81.6 (1.2)\\
\addlinespace[2pt]\multicolumn{9}{@{}l}{\textbf{Linear informative censoring, 30\% censored}}\\*
1\% & 0.2928 & CIPCW (GAM) & $-$2.9 (0.3) & 11.3 (0.2) & $-$0.0085 (0.0010) & 0.0320 (0.0007) & 0.0331 (0.0007) & 84.8 (1.1)\\
10\% & 0.2439 & CIPCW (GAM) & $-$3.6 (0.4) & 14.3 (0.3) & $-$0.0089 (0.0011) & 0.0337 (0.0008) & 0.0348 (0.0007) & 85.3 (1.1)\\
25\% & 0.1886 & CIPCW (GAM) & $-$4.6 (0.6) & 19.5 (0.4) & $-$0.0087 (0.0011) & 0.0358 (0.0008) & 0.0368 (0.0008) & 84.8 (1.1)\\
50\% & 0.1248 & CIPCW (GAM) & $-$4.7 (1.1) & 34.6 (0.9) & $-$0.0059 (0.0014) & 0.0428 (0.0010) & 0.0432 (0.0011) & 88.9 (1.0)\\
75\% & 0.0727 & CIPCW (GAM) & $-$1.8 (2.7) & 83.9 (3.9) & $-$0.0013 (0.0019) & 0.0610 (0.0014) & 0.0610 (0.0028) & 86.5 (1.1)\\
\end{longtable}
}

\clearpage
\section{GBSG case study: censoring model}\label{sec:gbsg}

\begin{table}[!htbp]
\centering\small
\caption{Cox proportional hazards model for the censoring times in the GBSG validation cohort ($n = 686$, 387 censored). \textit{Abbreviations:} CI = confidence interval, GBSG = German Breast Cancer Study Group.}\label{tab:gbsg}
\begin{tabular}{@{}l c c c@{}}
\toprule
 & Hazard ratio (95\% CI) & $p$ & Schoenfeld test $p$ \\
\midrule
\multicolumn{4}{@{}l}{\textit{Model coefficients}}\\
Age (per year) & 0.992 (0.975--1.009) & 0.333 & 0.036 \\
Postmenopausal (vs premenopausal)$^a$ & 0.853 (0.603--1.206) & 0.368 & 0.411 \\
Tumor size (per mm) & 1.003 (0.996--1.011) & 0.389 & 0.779 \\
Positive lymph nodes (per node) & 1.019 (0.995--1.044) & 0.129 & 0.569 \\
Estrogen receptor (per 100 fmol/L) & 1.037 (0.961--1.119) & 0.345 & 0.529 \\
Hormonal treatment (yes vs no)$^a$ & 0.526 (0.363--0.762) & 0.001 & 0.922 \\
Hormonal treatment $\times$ postmenopausal & 1.884 (1.190--2.982) & 0.007 & 0.136 \\
Global test & & & 0.080 \\
\midrule
\multicolumn{4}{@{}l}{\textit{Hormonal treatment (yes vs no), by menopausal status}}\\
Premenopausal & 0.526 (0.363--0.762) & 0.001 & \\
Postmenopausal & 0.991 (0.755--1.299) & 0.945 & \\
\bottomrule
\end{tabular}

\vspace{0.3em}
\parbox{0.9\textwidth}{\footnotesize $^a$Because the model includes the interaction between hormonal treatment and menopausal status, the hormonal treatment coefficient is the hazard ratio in premenopausal women and the menopausal status coefficient is the hazard ratio in women without hormonal treatment. The hazard ratio in postmenopausal women (bottom block) combines the hormonal treatment and interaction coefficients. The global Schoenfeld residual test did not indicate a violation of the proportional hazards assumption, although the test for age suggested some departure.}
\end{table}

\clearpage
\section{Positivity diagnostic}\label{sec:positivity}

\begin{table}[!htbp]
\centering\small
\caption{Probability of remaining uncensored at 5 years by decision threshold, standard survival setting with linear informative censoring, computed in the reference population of one million individuals using the true censoring distribution. Individuals above the threshold are those classified as high risk by the prediction model. Probabilities below 0.02 were truncated at 0.02 when the censoring weights were estimated. \textit{Abbreviation:} $G(5\mid\mathbf{X})$ = probability of remaining uncensored at 5 years given the covariates.}\label{tab:positivity}
\begin{tabular}{@{}r r r r r r@{}}
\toprule
 & & \multicolumn{3}{c}{Above threshold} & Below threshold\\
\cmidrule(lr){3-5}\cmidrule(lr){6-6}
Censoring & Threshold & \% of population & Median $G(5\mid\mathbf{X})$ & \% below 0.02 & Median $G(5\mid\mathbf{X})$\\
\midrule
10\% & 0.01 & 94.6 & 0.915 & 1.1 & 0.997\\
10\% & 0.10 & 62.2 & 0.833 & 1.7 & 0.982\\
10\% & 0.25 & 40.4 & 0.721 & 2.6 & 0.968\\
10\% & 0.50 & 23.3 & 0.552 & 4.5 & 0.953\\
10\% & 0.75 & 13.3 & 0.365 & 7.9 & 0.942\\
20\% & 0.01 & 95.1 & 0.811 & 3.8 & 0.993\\
20\% & 0.10 & 62.8 & 0.652 & 5.8 & 0.959\\
20\% & 0.25 & 40.6 & 0.463 & 9.0 & 0.926\\
20\% & 0.50 & 23.2 & 0.242 & 15.7 & 0.892\\
20\% & 0.75 & 13.0 & 0.088 & 28.0 & 0.867\\
30\% & 0.01 & 95.5 & 0.687 & 7.8 & 0.988\\
30\% & 0.10 & 64.0 & 0.471 & 11.7 & 0.930\\
30\% & 0.25 & 41.6 & 0.258 & 17.9 & 0.873\\
30\% & 0.50 & 23.8 & 0.082 & 31.3 & 0.815\\
30\% & 0.75 & 13.4 & 0.013 & 55.8 & 0.774\\
\bottomrule
\end{tabular}
\end{table}

\clearpage
\section{Consistency and asymptotic unbiasedness of the CIPCW-based estimators}\label{sec:proof}

We use Assumptions (1)--(3) of the main text, with (3) holding uniformly, $D = \sup_{u \le \tau,\, x} |\widehat{G}(u \mid x) - G(u \mid x)| \xrightarrow{p} 0$, and assume $0 < \mathrm{Prev}_\tau^{(k)} < 1$ and $P(T = C) = 0$. The function $\pi^{(k)}(\cdot, \tau)$ is fixed because the model is developed in independent data. For $u \le \tau$, $G(u \mid x) \ge G(\tau \mid x) \ge \epsilon$.

\textbf{Lemma S1.} For each cause $j \in \{1, \ldots, M\}$,
\begin{align}
E\left\{ \frac{I(\widetilde{T}_i \le \tau, \Delta_i = j)}{G(\widetilde{T}_i \mid X_i)} \,\middle|\, T_i, J_i, X_i \right\} &= I(T_i \le \tau, J_i = j), \tag{L1}\\
E\left\{ \frac{I(\widetilde{T}_i > \tau)}{G(\tau \mid X_i)} \,\middle|\, T_i, J_i, X_i \right\} &= I(T_i > \tau). \tag{L2}
\end{align}
Hence, (L1) with $j = k$ gives (L3), and (L2) plus the sum of (L1) over $j \notin \{0, k\}$ gives (L4):
\begin{align}
E\left\{ \frac{I(\widetilde{T}_i \le \tau, \Delta_i = k)}{G(\widetilde{T}_i \mid X_i)} \,\middle|\, T_i, J_i, X_i \right\} &= I(T_i \le \tau, J_i = k), \tag{L3}\\
E\left\{ \frac{I(\widetilde{T}_i > \tau)}{G(\tau \mid X_i)} + \frac{I(\widetilde{T}_i \le \tau, \Delta_i \notin\{0,k\})}{G(\widetilde{T}_i \mid X_i)} \,\middle|\, T_i, J_i, X_i \right\} &= I(T_i > \tau) + I(T_i \le \tau, J_i \ne k). \tag{L4}
\end{align}

\textit{Proof.} For $j \ne 0$, $I(\widetilde{T}_i \le \tau, \Delta_i = j) = I(T_i \le \tau, J_i = j)\, I(C_i \ge T_i)$, and $\widetilde{T}_i = T_i$ when this is non-zero. Since $P(T = C) = 0$, $I(C_i \ge T_i) = I(C_i > T_i)$ almost surely, so
\[
E\left\{ \frac{I(\widetilde{T}_i \le \tau, \Delta_i = j)}{G(\widetilde{T}_i \mid X_i)} \,\middle|\, T_i, J_i, X_i \right\}
= \frac{I(T_i \le \tau, J_i = j)}{G(T_i \mid X_i)}\, P(C_i > T_i \mid T_i, J_i, X_i)
= I(T_i \le \tau, J_i = j),
\]
because, by Assumption (1), $P(C_i > T_i \mid T_i, J_i, X_i) = G(T_i \mid X_i)$. Similarly, $I(\widetilde{T}_i > \tau) = I(T_i > \tau)\, I(C_i > \tau)$ and $P(C_i > \tau \mid T_i, J_i, X_i) = G(\tau \mid X_i)$ give (L2). \hfill $\square$

\textbf{Theorem S1.} $\widehat{\mathrm{Prev}}_\tau^{(k)}$, $\widehat{\mathrm{Se}}_\tau^{(k)}$, $\widehat{\mathrm{Sp}}_\tau^{(k)}$ and $\widehat{\mathrm{NB}}_\tau^{(k)}$ converge in probability to $\mathrm{Prev}_\tau^{(k)}$, $\mathrm{Se}_\tau^{(k)}$, $\mathrm{Sp}_\tau^{(k)}$ and $\mathrm{NB}_\tau^{(k)}$, and their expectations converge to the same limits.

\textit{Proof.} The numerator of $\widehat{\mathrm{Prev}}_\tau^{(k)}$ is the denominator of $\widehat{\mathrm{Se}}_\tau^{(k)}$, and its denominator is the sum of the denominators of $\widehat{\mathrm{Se}}_\tau^{(k)}$ and $\widehat{\mathrm{Sp}}_\tau^{(k)}$. Each estimator is therefore a ratio of averages of the form $n^{-1}\sum_i$ of the summands in Equation (3).

For the true weights, that is, with $G$ in place of $\widehat{G}$, the summands are i.i.d.\ and lie in $[0, 1/\epsilon]$, so each average converges in probability to the expectation of one summand. Since $I(\pi_i^{(k)} > z)$ depends only on $X_i$, iterated expectation and (L3) give
\[
E\left\{ \frac{I(\pi_i^{(k)} > z, \widetilde{T}_i \le \tau, \Delta_i = k)}{G(\widetilde{T}_i \mid X_i)} \right\}
= P(\pi^{(k)} > z, T \le \tau, J = k) = \mathrm{Prev}_\tau^{(k)}\, \mathrm{Se}_\tau^{(k)} .
\]
In the same way, (L3) and (L4) give the limits $\mathrm{Prev}_\tau^{(k)}$, $(1 - \mathrm{Prev}_\tau^{(k)})\,\mathrm{Sp}_\tau^{(k)}$ and $1 - \mathrm{Prev}_\tau^{(k)}$ for the other three averages, and hence 1 for the denominator of $\widehat{\mathrm{Prev}}_\tau^{(k)}$.

For the estimated weights, when $D < \epsilon/2$, $\widehat{G}(u \mid x) \ge \epsilon/2$ for all $u \le \tau$ and $x$, so
\[
\left| \frac{1}{\widehat{G}(u \mid x)} - \frac{1}{G(u \mid x)} \right| \le \frac{2D}{\epsilon^2} .
\]
Each summand contains a single weight evaluated at a time no later than $\tau$, so each average computed with $\widehat{G}$ differs from the one computed with $G$ by at most $2D/\epsilon^2$. Since $D \xrightarrow{p} 0$, the averages have the same limits. This bound needs no independence between $\widehat{G}$ and the summands, so $\widehat{G}$ may be fitted on the same data.

For consistency, the limits of the denominators are positive because $0 < \mathrm{Prev}_\tau^{(k)} < 1$, and Equation (2) is continuous in $(\mathrm{Prev}_\tau^{(k)}, \mathrm{Se}_\tau^{(k)}, \mathrm{Sp}_\tau^{(k)})$, so the continuous mapping theorem gives convergence in probability of all four estimators.

For asymptotic unbiasedness, in each of $\widehat{\mathrm{Prev}}_\tau^{(k)}$, $\widehat{\mathrm{Se}}_\tau^{(k)}$ and $\widehat{\mathrm{Sp}}_\tau^{(k)}$ the numerator is part of the denominator, so each lies in $[0, 1]$, and $\widehat{\mathrm{NB}}_\tau^{(k)}$ lies in $[-z/(1-z), 1]$. Bounded sequences that converge in probability also converge in mean, so the expectations converge to the same limits. \hfill $\square$

\end{document}